\documentclass[a4paper,11pt]{article}
\usepackage[left=2cm,right=2cm,top=2cm,bottom=2cm]{geometry}
\usepackage{amsmath,amsfonts,amssymb,amsthm,verbatim,lscape}
\usepackage{commath}
\usepackage[usenames,dvipsnames,svgnames,table]{xcolor}
\usepackage[figuresright]{rotating}

\usepackage{hyperref}
\hypersetup{colorlinks,
citecolor=black,
filecolor=black,
linkcolor=black,
urlcolor=black,
pdftex}
\usepackage{color}

\usepackage{graphicx}

\usepackage{longtable}
\usepackage{setspace}

\usepackage[font={footnotesize,it}]{caption}

\usepackage[compress]{natbib}
\begin{document}
\def\a{\alpha}
\newcommand{\bea}{\begin{eqnarray}}
\newcommand{\eea}{\end{eqnarray}}

\def\th{\theta}
\def\g{\gamma}
\def\thbf{\boldsymbol{\theta}}
\def\dbf{\boldsymbol{\delta}}

\def\R{\mathbf{R}}
\def\x{\mathbf{x}}
\def\X{\mathbf{X}}
\def\Y{\mathbf{Y}}
\def\y{\mathbf{y}}
\def\a{\alpha}
\def\s{\mathbf{s}}
\def\R{\mathbf{R}}
\def\I{\mathbf{I}}
\def\x{\mathbf{x}}
\def\X{\mathbf{X}}
\def\bfv{\mathbf{v}}
\def\V{\mathbf{V}}
\def\Y{\mathbf{Y}}
\def\y{\mathbf{y}}
\def\bbf{\boldsymbol{\beta}}
\def\gbf{\boldsymbol{\gamma}}
\def\obf{\boldsymbol{\omega}}
\def\Obf{\boldsymbol{\Omega}}
\def\U{\mathbf{U}}
\def\J{\mathbf{J}}
\def\ga{\gamma}
\def\u{\mathbf{u}}
\def\C{\mathbf{C}}
\def\p{\mathbf{p}}
\def\1{\mathbf{1}}
\def\t{\mathbf{t}}
\def\W{\mathbf{W}}
\def\Cbar{{\overline C}}
\def\C{\mathbf{C}}

\newcommand{\widesim}[2][1.5]{
  \mathrel{\overset{#2}{\scalebox{#1}[1]{$\sim$}}}
}

\def \red#1{\textcolor{red}{#1}}
\def \blue#1{\textcolor{blue}{#1}}
\def \green#1{\textcolor{teal}{#1}}

\hyphenation{nik-ol-oul-op-oul-os}
\title{A  copula-based model for multivariate  ordinal panel data:\\ application to well-being composition}
\date{}

\author{
Aristidis K. Nikoloulopoulos\footnote{{\small\texttt{A.Nikoloulopoulos@uea.ac.uk}}, School of Computing Sciences, University of East Anglia,
Norwich NR4 7TJ, UK} 
\and  Emmanouil Mentzakis\footnote{{\small\texttt{E.Mentzakis@soton.ac.uk}}, Economics Department, School of Social Sciences, University of Southampton, Southampton SO17 1BJ, UK}
}
\maketitle

\begin{abstract}
\baselineskip=18pt
\noindent  
A novel copula-based multivariate panel ordinal model is developed to  estimate 
structural relations among components of well-being.  
Each ordinal  time-series is modelled  using a copula-based Markov model to relate the marginal distributions of the response at each time of observation and then, at each observation time, the conditional  distributions of each ordinal time-series are joined using a multivariate t copula.
Maximum simulated likelihood based on evaluating the multidimensional integrals of the likelihood with randomized quasi Monte Carlo methods is used for the estimation. Asymptotic calculations show that our method is nearly as efficient as maximum likelihood for fully specified multivariate copula models. 
Our findings highlight the importance of one's relative position in evaluating their well-being with no direct effects of socio-economic characteristics on well-being but strong indirect effects through their impact on components of well-being. Temporal resilience, habit formation and behavioural traits can explain the  dependence in the joint tails over time and across well-being components.


\end{abstract}

\baselineskip=20pt
\section{Introduction}

Subjective well-being and life satisfaction have received a significant amount of attention in the economics  \citep{blanchflower_well-being_2004, ferrer-i-carbonell_how_2004} and psychology \citep{diener_subjective_1999, argyle_causes_1999} literature over the past decades. While cardinality properties are conceptually rejected, the ordinal nature of the information conveyed in such indicators has gradually entered public policy circles. \citet{stiglitz_report_2009} argued that meaningful and reliable data on well-being can be collected and should be included in surveys and official statistics with further reports advocating its role in policy making in informing policy designs, monitoring progress and evaluating implementation of public interventions \citep{dolan_measuring_2012}.

However, despite suggestions by various authors \citep{cummins_developing_2003, kahneman_developments_2006, krueger_reliability_2008, veenhoven_four_2013}, there is little consensus in how exactly such concepts are defined   
and what precisely they capture. For the past few years the Office for National Statistics in the UK monitors year-on-year improvements in reported well-being through questions on life satisfaction, feelings of self-worth, happiness and anxiety, which however provides little theoretical basis or conceptual background for such choices \citep{ons_personal_2014}. While it is clear that well-being is less of an answer to a single question and more of a composite multidimensional concept, limited  empirical work has pursued such considerations. The vast majority of empirical work (for reviews see \citet{kahneman_developments_2006,dolan_we_2008}) focuses on univariate models that would not allow taking into account dependence between constituent components and their link with overall happiness, while also being unable to correctly identify direct effects of independent determinants/variables on domain and generic happiness in the presence of such outcomes' dependence. In our context, generic well-being or happiness pertains to one's satisfaction with their life overall, while domain well-being/satisfaction pertains to satisfaction individual constituent components of life, e.g. income, health, social life, family life, etc.

Conceptual work suggests bottom-up theories \citep{diener_subjective_1984} for life-satisfaction where judgements are based on assessments of satisfaction with a number of defined life domains with ensuing causal pathways running from the domains upwards \citep{schimmack_structure_2008}. Complementary positions view overall life satisfaction as the net outcome of reported satisfaction with life domains with the domains themselves seen as functions of objective outcomes/situations/covariates \citep{michalos_global_1991, easterlin_happiness_2008}. Looking at the structure of composite well-being, \citet{salvatore_appraisal_2001} conclude it is more easily interpreted in terms of generic dimensions of life (e.g. family life, social life, love life, occupational life and leisure) than attributed to the fulfilment of personal values (self-acceptance, autonomy, environmental mastery, purpose in life, and personal growth). \citet{brief_integrating_1993} argued for a mixture of processes where individual characteristics and traits are driving domain satisfaction (i.e. top-down), which in turn drives life satisfaction (i.e. bottom-up). In other words, high individual income leads to increased life satisfaction because financial satisfaction is an important component of satisfaction with life as a whole. 

Looking at the economics of well-being literature, univariate models again dominate, while at the same time very few studies have focused on its composite structure. \citet{van_praag_anatomy_2003} formulate a model very close to the initial bottom-up theories, which however is estimable only under strong assumptions of no direct association between individual covariates and generic life satisfaction. However, problems of omitted variables and lack of sufficient exclusion restrictions suggest caution in drawing conclusions. Similar problems are also faced in \citet{easterlin_happiness_2008}, where again strong assumptions (both conceptual and econometric) are required in the model and raise concerns about the robustness of findings.

However, most theories developed within the well-being composition literature offer limited empirical evidence for the underlying connections posited and fail to explicitly incorporate them into their setting. In this paper we propose a multivariate framework and estimate a comprehensive relationships pattern between generic and domain satisfactions with dependence explicitly modelled through copulas. We develop a novel joint copula-based Markov model, where  a set of bivariate copulas and a multivariate t (MVT) copula  jointly model multivariate ordinal time-series responses with covariates. Each ordinal time-series is  considered a copula-based Markov model, where a parametric bivariate copula family is used for the joint distribution of subsequent observations and which is then related to these ordinal time-series responses using an MVT to join their conditional (on the past) distributions at each time point. Note in passing that  other continuous-variable models  using copulas in several Markov chains  exist in the literature \citep{lambert&vande02,Patton-2012,Remillard-etal-2012, Beare-etal-2015}, but in our knowledge we are the first constructing  such copula-based models for ordinal time-series with covariates. The theoretical and estimation   concepts in the discrete case are quite different.   

Simple parametric families of copulas in more than two dimensions  typically  provide limited dependence \citep{Nikoloulopoulos2013a} and for discrete data it is generally hard to provide a better fit than the multivariate normal (MVN) copula, which inherits  the useful properties of the MVN distribution. 
However, the MVN copula is inadequate to model multivariate data with  more probability in one or both joint tails. 
In recent years, a popular and useful approach is the vine pair-copula construction \citep{Kurowicka-Joe-2011,nikoloulopoulos&joe&li11} which is based on $d(d-1)/2$ bivariate copulas, of which some are used to summarize conditional dependence. Vine copula constructions are suitable for modelling this kind of data  by using appropriate  bivariate copulas.   \cite{panagiotelis&czado&joe12} and \cite{nikoloulopoulos&joe12} recently extended the idea of vine copulas to discrete data. The approach in \cite{nikoloulopoulos&joe12} involves both observed and latent variables, while the approach of \cite{panagiotelis&czado&joe12} is suitable when there are no latent variables to explain the dependence in the observed variables. In this paper to form the bivariate part of the model,  a D-vine truncated at the 1-st level \citep{Brechmann-Czado-Aas-2012} has been exploited.

To develop the joint copula-based Markov model we propose the  MVT copulas. 
MVT copulas nest MVN copulas 
and share with them the ability to accommodate any feasible pattern of association in a set of random variables. However, 
 the MVT copulas offer greater flexibility than MVN copulas, as they can also capture dependence in both joint tails \citep{Nikoloulopoulos&joe&li09}, which is  the case in `mixtures' of population (e.g., different locations or genders). The MVT copula-based approach of this article avoids having to specify parametrically the distribution of latent heterogeneity in a non-linear setting. The MVT copulas  as  scale mixtures of MVN can be used to explore unobserved population heterogeneity.

Implementation of the MVT copula for discrete data is feasible, but not easy, because the MVT distribution, as a latent model for discrete response, requires rectangle probabilities based on high-dimensional integrations \citep{genest&nikoloulopoulos&rivest08}. 
The probability mass function (pmf) can be  
obtained by computing  an MVT rectangle probability and the randomized quasi Monte Carlo methods proposed by \cite{genz&bretz02} can be used for that purpose. Computing the rectangle MVT probabilities via simulation based on the methods in \cite{genz&bretz02} is akin to using a simulated likelihood method, whose asymptotic efficiency, for the special case of MVN copula, has been studied by \citet{nikoloulopoulos13b,
nikoloulopoulos2015a} and was shown to be as good as maximum likelihood for dimension 10 or lower. 

Modelling of dependence further allows revisiting a number of established relationships in the literature between covariates and well-being and examining their 
association in a multivariate setting (e.g. does the importance of income on generic well-being remain when considered along income satisfaction?). In short, we establish exogenous determinants for both generic and domain satisfaction equations and separately identify structural relations among components of well-being and shed light not only into their links with generic well-being but also among themselves. This allows separate identification of the direct effect of domain characteristics and of composite domains on generic satisfaction, as well as estimation of the dependence between and among generic and domain satisfaction.

\cite{Prokhorov&Schmidt2009} showed that robust estimation can be achieved in the class of radially or reflection symmetric copulas such as the normal or t copula. \cite{nikoloulopoulos&joe&chaganty10}, \cite{Masarotto&Varin12}, \cite{Nikoloulopoulos2015d} and \cite{Nikoloulopoulos-2016-wtsc-ord} (in particular for ordinal time-series) showed robustness of the normal copula  to dependence if the main interest is the univariate parameters (regression and nonregression parameters). 
This type of research focussed on marginal  models and on estimation of coefficients for regression models with time-series data, that are robust to the dependence structure. 
Nevertheless, our manuscript is not  entirely in the area of ``marginal models" (meaning specification of univariate time-series only), but also avails copula-based models  for both univariate and multivariate time-series.
 One of the  goals of our paper is to compare dependence models in inferences involving joint probabilities. 

The remainder of the paper proceeds as follows.  Section \ref{themodel} introduces the joint copula-based Markov model for discrete ordinal responses with covariates and   presents the conceptual framework upon which this econometric model is built. Estimation techniques and computational details are provided in Section \ref{estimation}.  
Section \ref{vuong-sec} discusses   the \nocite{vuong1989}Vuong's  (1989) test to assess the fit of the proposed model in terms of prediction of joint probabilities. Section \ref{sec-appl} presents the application of our methodology to the British Household Panel and Section \ref{sec-discussion} concludes, followed by a technical Appendix.

\section{\label{themodel}A joint copula-based Markov model}
In this section, we construct the joint copula-based Markov model for ordinal time-series with covariates. Before that, the first subsection has some background on copula models.

\subsection{\label{overview}Overview and relevant background for copulas}
A copula is a multivariate cdf with uniform $U(0,1)$ margins \citep{joe97,joe2014,nelsen06}.
If $F$ is a $d$-variate cdf with univariate margins $F_1,\ldots,F_d$,
then Sklar's (1959) theorem\nocite{sklar1959} implies that there is a copula $C$ such that
  $$F(y_1,\ldots,y_d)= C\Bigl(F_1(y_1),\ldots,F_d(y_d)\Bigr).$$
The copula is unique if $F_1,\ldots,F_d$ are continuous, but not
if some of the $F_j$ have discrete components.
If $F$ is continuous and $(Y_1,\ldots,Y_d)\sim F$, then the unique copula
is the distribution of $(U_1,\ldots,U_d)=\left(F_1(Y_1),\ldots,F_d(Y_d)\right)$ leading to
  $$C(u_1,\ldots,u_d)=F\Bigl(F_1^{-1}(u_1),\ldots,F_d^{-1}(u_d)\Bigr),
  \hspace{2ex} 0\le u_j\le 1, j=1,\ldots,d,$$
where $F_j^{-1}$ are inverse cdfs. In particular,
if $\mathcal{T}_d(\cdot;\R)$
is the MVT cdf with correlation  matrix $\R=(\rho_{jk}: 1\le j<k\le d)$ and $\nu$  degrees of freedom, and $\mathcal{T}$ is the univariate Student t cdf with $\nu$  degrees of freedom,
then the MVT copula is
\begin{equation}\label{MVNcdf}
C(u_1,\ldots,u_d)=\mathcal{T}_d\Bigl(\mathcal{T}^{-1}(u_1),\ldots,\mathcal{T}^{-1}(u_d);\R\Bigr).
\end{equation}

\subsection{\label{markov}Copula-based Markov models  for ordinal time-series }

For a latent variable $Z\sim \mathcal{F}$ such that $Y=y$ if
$\alpha_{y-1}+\x^T\bbf\leq Z\leq  \alpha_{y}+\x^T\bbf,\,y=1,\ldots,K,$
with $K$ being the number of categories of $Y$, $\bbf$ the $p$-dimensional regression vector, $t=1 \dots T$ the ``panel'' dimension, $i=1 \dots n$ the number of clusters (note that varying cluster sizes can be accommodated by the theory) and $p$ the number of covariates (i.e. the dimension of a covariate vector $\x$), the response $Y$ is assumed to have density
$$f(y;\mu,\gbf)=\mathcal{F}(\alpha_{y}+\mu)-\mathcal{F}(\alpha_{y-1}+\mu),$$
where $\mu=\x^T\bbf$ is a function of $\x$, $\bbf$ is a $p$-dimensional regression vector and $\gbf=(\alpha_1,\ldots,\alpha_{K-1})$ is the $q$-dimensional vector of the univariate cutpoints ($q=K-1$) with  $\alpha_0=-\infty$ and $\alpha_K=\infty$. Choosing normal or logistic for $\mathcal{F}$ leads to the ordered probit and cumulative logit models, respectively. 

For data $(y_{itj}, \x_{itj})$, where $j$ is an index for the ordinal responses, the univariate marginal model for
$Y_{itj} $ is $f_j(y_{itj}; \mu_{itj},\gbf_j)$ where  $\mu_{itj}=\x_{itj}^\top\bbf_j$ and  $\gbf_j$ of dimension $q_j$. If for each $t$, $Y_{i1j},\ldots,Y_{iTj}$ are independent, then the
log-likelihood for each univariate ordinal response is
\begin{eqnarray}\label{indlik}\ell_{j}= \sum_{i=1}^n\sum_{t=1}^T\, \log f_j(y_{itj};\mu_{itj},\gbf_j).
\end{eqnarray}

If the ordinal data are observed in a time-series sequence, then the ordinal regression model can be adapted in two ways: 
\begin{enumerate}
\item[1.] add previous observations as covariates;
\item[2.] make use of some models for stationary ordinal time-series.  
\end{enumerate}  
Here we adapt the methodology for case 2.  
For dependent $Y_{i1j},\ldots,Y_{iTj}$, estimation of $\bbf_j$ and $\gbf_j$ involves copula-based Markov models \citep[page 244]{joe97} for ordinal time-series with covariates. The joint distribution of subsequent observations is modelled through a parametric copula family with the corresponding transition probabilities subsequently elicited. 
The advantages of a time-series regression model are explicitly mentioned in \cite{Joe-2015-proceedings} and also reproduced below:

\begin{itemize}
\itemsep=0pt

\item The class of autocorrelation functions is much wider than those based on an ordered probit with lagged dependent variables appearing as explanatory variables.

\item Prediction in regressions with time dependent observations is simpler as they can be formulated with or without the preceding observations.

\item Serial dependence (positive or negative) can be modelled through suitable copula families.

\item The non-linearity of the conditional expectations allows for various patterns to be replicated, while the conditional expectation and variance for large values is determined by the choice of copula family and corresponding tail behaviour.

\item Incorporating covariates in time-series models is more straightforward in univariate regression models.

\item Extending the framework to Markov orders higher than one is straightforward.

\item Copula families with an easy (e.g. closed) form allow for easier likelihood inference.

\end{itemize}
Note in passing that copula-based Markov models for continuous response data have been studied before in \citet{chen-fan-06}. 

Assuming a copula based Markov model, the transition cdf of $Y_{tj}$ given $Y_{t-1,j}$ is
\begin{eqnarray}\label{transition-cdf}F_{j|t}(y_{tj}|y_{t-1,j})&=&P(Y_{tj}\leq
y_{tj}|Y_{t-1,j}=y_{t-1,j})\\&=&\Bigl[C_{j|t}\bigl(F(y_{t-1,j}),F(y_{tj})\bigr)-C_{j|t}\bigl(F(y_{t-1,j}-1),F(y_{tj})\bigr)\Bigr]/f_j(y_{t-1,j}),\nonumber
\end{eqnarray}
and the transition pmf is
$$f_{j|t}(y_{tj}|y_{t-1,j})=P(Y_{tj}=y_{tj}|Y_{t-1,j}=y_{t-1,j})=\frac{f(y_{tj},y_{t-1,j})}{f_j(y_{t-1,j})},$$
where
$f(y_t,y_{t-1})=C_{j|t}\bigl(F(y_t),F(y_{t-1})\bigr)-C_{j|t}\bigl(F(y_t-1),F(y_{t-1})\bigr)-C_{j|t}\bigl(F(y_t),F(y_{t-1}-1)\bigr)+C_{j|t}\bigl(F(y_t-1),F(y_{t-1}-1)\bigr)$.
Then the log-likelihood for each ordinal time-series is 
\begin{equation}\label{serlik}\ell_{j|t}= \sum_{i=1}^n\left(\log f_j(y_{i1j};\mu_{i1j},\gbf_j)+\sum_{t=2}^T\, \log f_{j|t}(y_{itj}|y_{i,t-1,j};\mu_{itj},\mu_{i,t-1,j},\gbf_j)\right).
\end{equation}

Such framework incorporates the BVN copula as a special case, i.e. ``autoregressive-to-anything'' in \cite{biller&nelson2005} as also acknowledged by \cite{joe2014}. Stronger clustering of consecutive large/small values than expected in the BVN would require alternative copulas to obtain more appropriate transition probabilities.

\subsubsection{\label{sec-families}Choices of parametric families of copulas}  In our candidate set, families that are in line with the conditions under which a copula function generates a stationary Markov chain that satisfies mixing conditions at a geometric rate \citep{chen-etal-2009,Beare2010} are used. These families 
have different strengths of tail behaviour (see e.g., \cite{nikoloulopoulos&joe&li11,
nikoloulopoulos&joe12}):
\footnote{A bivariate copula $C$ is {\it reflection symmetric}
if its density $c(u_1,u_2)=\partial^2 C(u_1,u_2)/\partial u_1\partial u_2$ 
 satisfies $c(u_1,u_2)=c(1-u_1,1-u_2)$ for all $0\leq u_1,u_2\leq 1$
and {\it reflection asymmetric} otherwise often with more probability in the
joint upper tail or joint lower tail. {\it Upper tail dependence} implies
 $c(1-u,1-u)=O(u^{-1})$ as $u\to 0$ and {\it lower tail dependence}
that $c(u,u)=O(u^{-1})$ as $u\to 0$.
If $(U_1,U_2)\sim C$ for a bivariate copula $C$, then $(1-U_1,1-U_2)\sim
C_{180^0}$, with $C_{180^0}(u_1,u_2)=u_1+u_2-1+C(1-u_1,1-u_2)$ being the survival (or rotated by 180 degrees) copula of $C$. The ``reflection"
of each uniform $U(0,1)$ random variable by about $1/2$ changes the direction
of tail asymmetry.}
\begin{itemize}
\itemsep=0pt
\item Frank copula is reflection symmetric satisfying tail independence 
$C(u,u)=O(u^2)$ and $\Cbar(1-u,1-u)=O(u^2)$ as $u\to 0$,
with cdf
$$C(u_1,u_2;\th)=-\theta^{-1}\log \left\{1+\frac{(e^{-\theta u_1}-1)(e^{-\theta
u_2}-1)}{e^{-\theta}-1} \right\},\hspace{2ex} \theta \in (-\infty,\infty)\setminus\{0\}.$$
\item Gumbel extreme value copula is reflection asymmetric with upper tail dependence and cdf
$$C(u_1,u_2;\th)=\exp\Bigl[-\Bigl\{(-\log u_1)^{\theta}
+(-\log u_2)^{\theta}\Bigr\}^{1/\theta}\Bigr],\hspace{2ex} \th\geq 1.$$
Compared to the BVN copula, the resulting model has more probability in the joint upper tail accommodating more dependence of  large ordinal values than expected in the BVN.

\item Survival Gumbel (s.Gumbel) copula is reflection asymmetric with  lower tail dependence and cdf
$$C(u_1,u_2;\th)=u_1+u_2-1 + \exp\Bigl[-\Bigl\{\bigl(-\log (1-u_1)\bigr)^{\theta}
+\bigl(-\log (1-u_2)\bigr)^{\theta}\Bigr\}^{1/\theta}\Bigr],\hspace{2ex} \th\geq 1.$$
Compared to the BVN copula, this resulting model has more probability in the joint lower tail accommodating more dependence of  small ordinal values than expected in the BVN. 
 
\item Bivariate Student t (BVT) copula with reflection symmetric upper and lower tail dependence and cdf
$$C(u_1,u_2;\th)=\mathcal{T}_2\Bigl(\mathcal{T}^{-1}(u_1;\nu),\mathcal{T}^{-1}(u_2;\nu);\th,\nu\Bigr),\hspace{2ex}-1\leq\th\leq 1,$$
where $\mathcal{T}(;\nu)$ is the univariate Student t cdf with (non-integer) $\nu$ degrees of freedom, and $\mathcal{T}_2$ is the
cdf of a bivariate Student t distribution with $\nu$ degrees of freedom and correlation parameter $\th$.
Small values of $\nu$ (i.e. $1\le \nu\le 5$) lead to models with more probabilities in the joint upper and joint lower tails accommodating more dependence of  large and small ordinal values that would be expected with BVN. 
\end{itemize}

\begin{figure}[!t]
\begin{center}
\begin{tabular}{cc}
\includegraphics[width=0.3\textwidth]{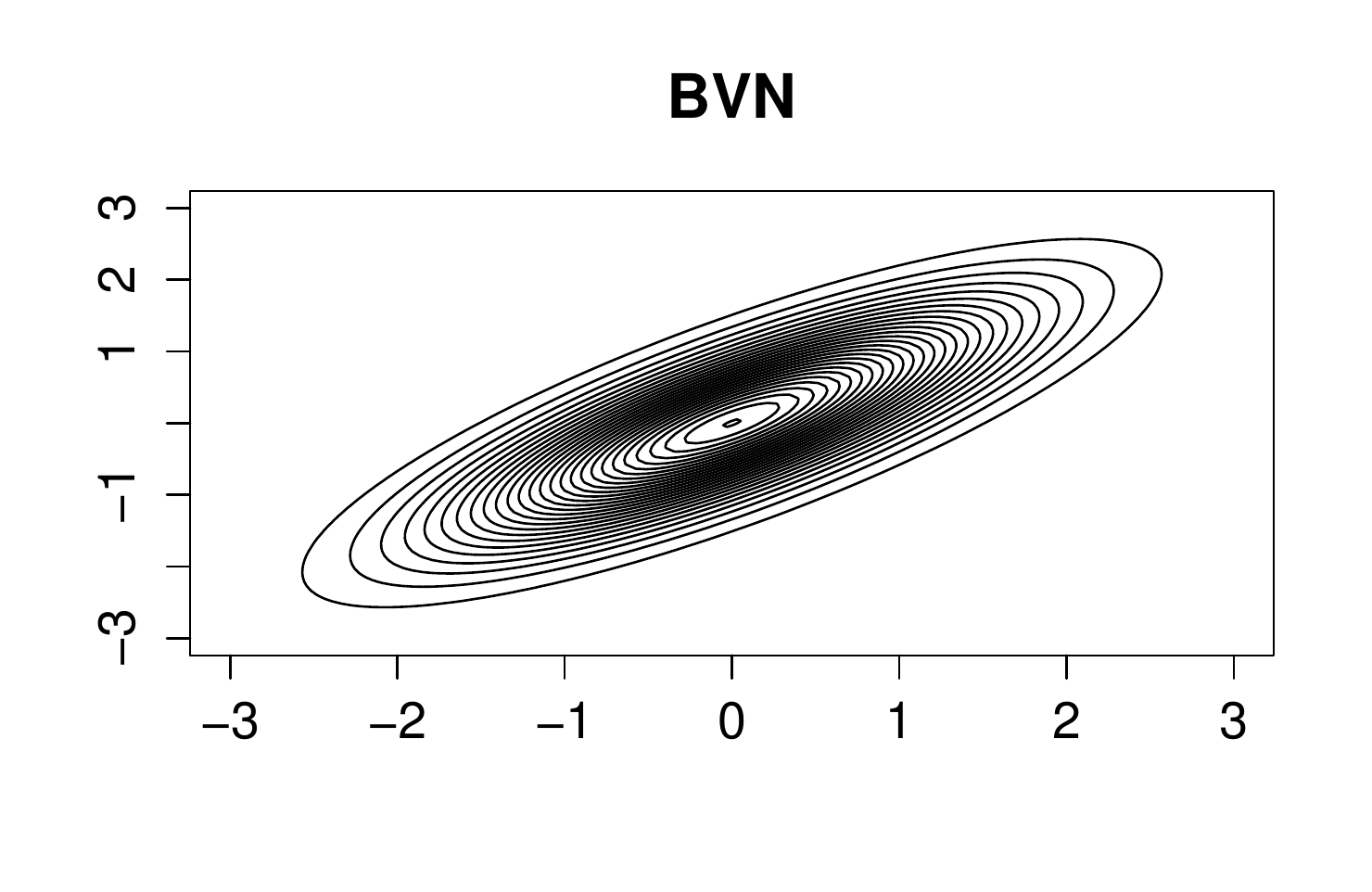}
&

\includegraphics[width=0.3\textwidth]{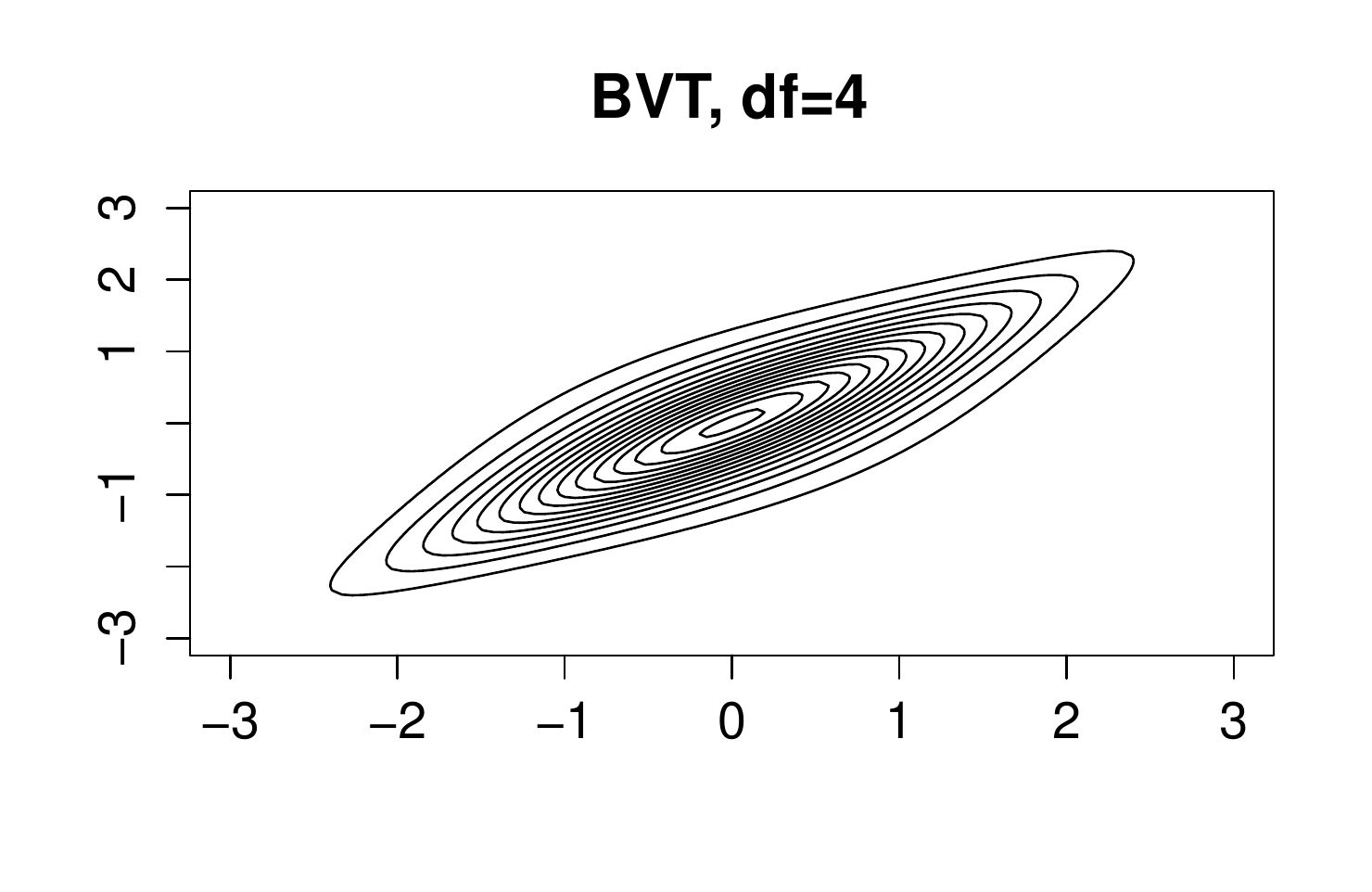}\vspace{-0.5cm}\\

\multicolumn{2}{c}{\includegraphics[width=0.3\textwidth]{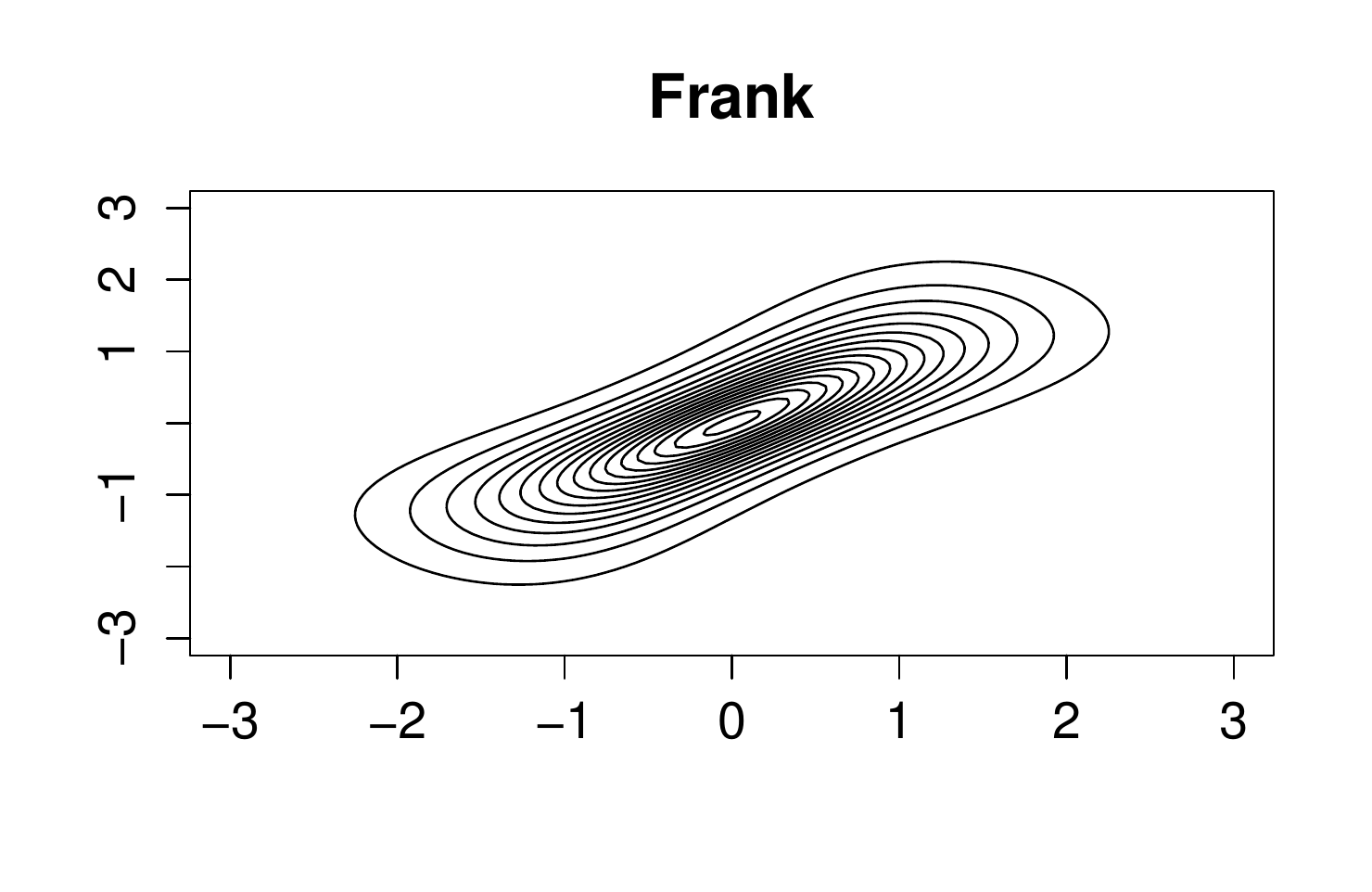}}\vspace{-0.5cm}
\\
\includegraphics[width=0.3\textwidth]{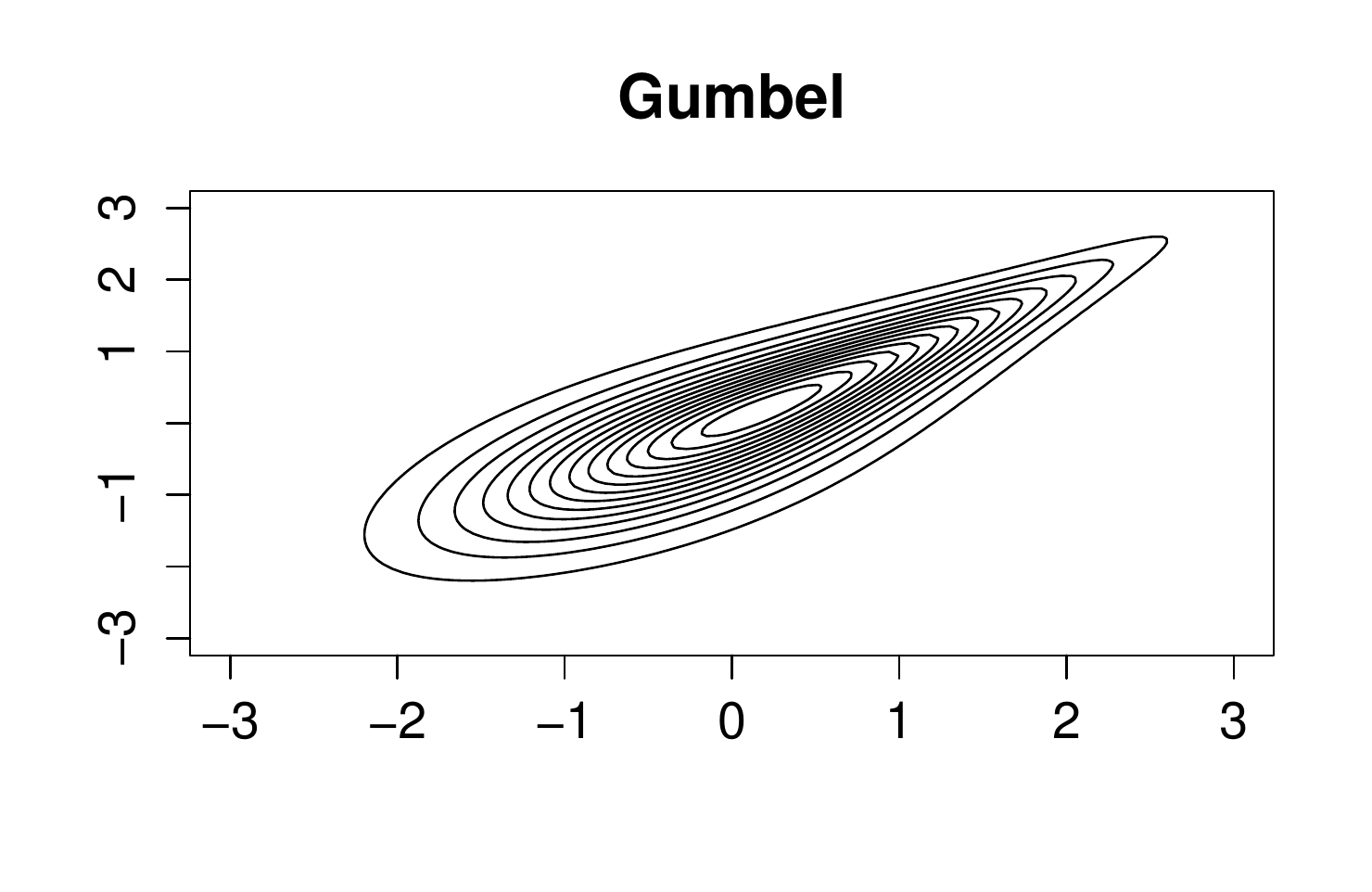}
&
\includegraphics[width=0.3\textwidth]{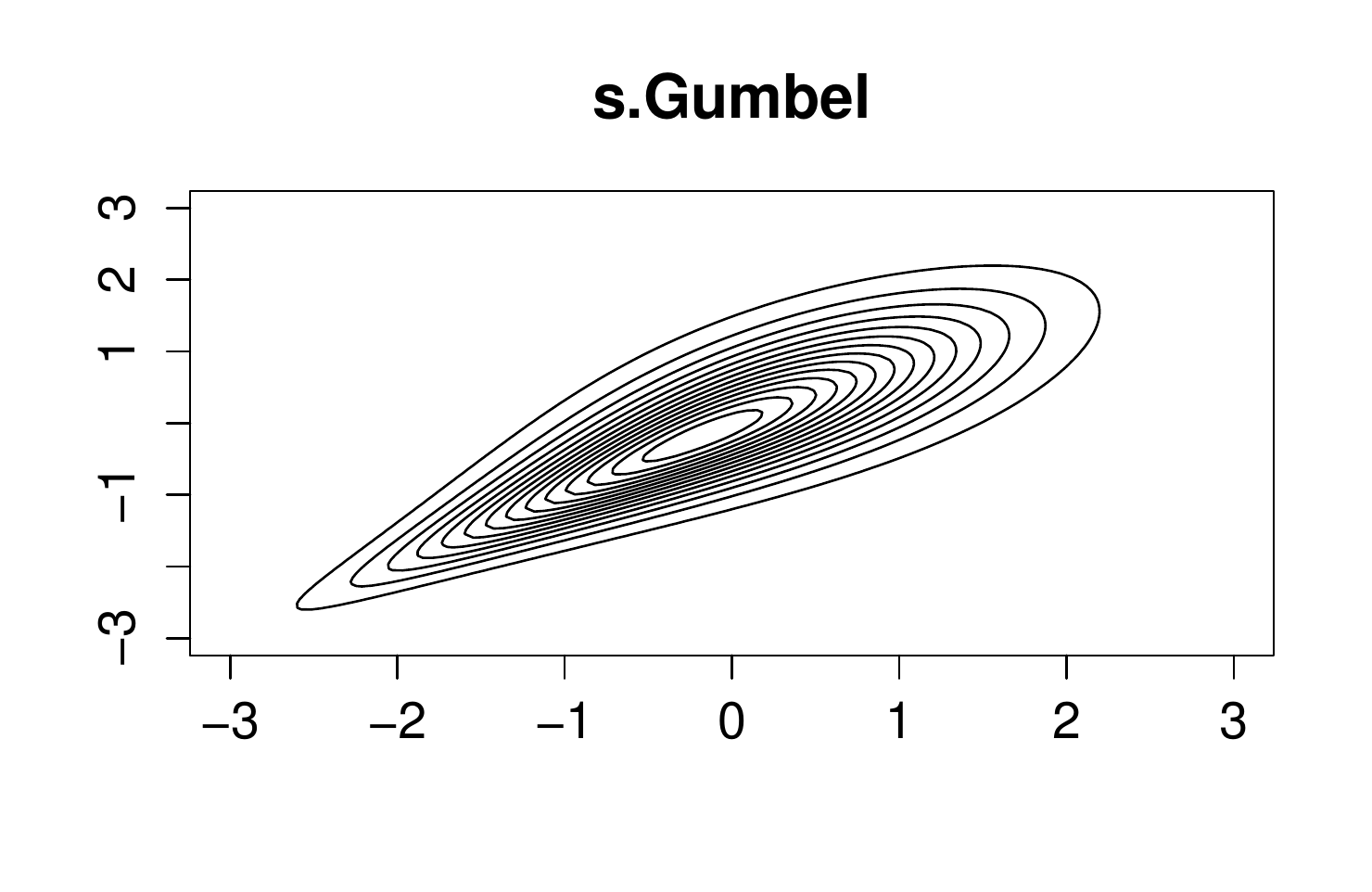}
\\
\end{tabular}
\caption{\label{contours}Contour plots of BVN, BVT with 4 degrees of freedom (df), Frank, Gumbel and s.Gumbel copulas with standard normal margins and dependence   parameters  corresponding to Kendall's $\tau$ value of $0.6$. }
\end{center}
\end{figure}

\citet{Nikoloulopoulos&karlis08CSDA} have shown that, when using real data, copulas with similar (tail) dependence properties provide similar fit making selection among them cumbersome. 
With tail dependence properties being copula family specific, upper/lower tail dependence becomes one way to differentiate among families. Contour plots of copula densities with standard normal margins and dependence parameters corresponding to Kendall's $\tau=0.6$ are given in Figure \ref{contours} to depict concepts of reflection (a)symmetric tail (in)dependence.

\subsection{Joint copula-based Markov models  for ordinal time-series}

So far we treat the $d$ ordinal time-series responses separately as if they were independent. In this Section, we propose relating these responses using an MVT copula to join their conditional (on the past) distributions at each time point.

Consider a multivariate discrete regression setup in which the $d \geq 2$ dependent ordinal time-series $Y_{t1}, \ldots, Y_{td}$ are observed together with a vector $\mathbf{x} \in \mathbb{R}^p$ of explanatory variables. If $C(\cdot;\R)$ is the MVT copula (or any other  parametric family of copulas) and $F_{j|t}(y_{tj}|y_{t-1,j})$, as defined in (\ref{transition-cdf}), 
is the   parametric model for the $j$th univariate ordinal  time-series  then   $$C\Bigl(F_{1|t}(y_{t1}|y_{t-1,1}),\ldots,F_{d|t}(y_{td}|y_{t-1,d});\R\Bigr)$$ is a multivariate parametric model with univariate margins $F_{1|t},\ldots,F_{d|t}$. For copula models, the response vector $\Y=(Y_1,\ldots,Y_d)$ can be  discrete \citep{Nikoloulopoulos2013a,nikoloulopoulos&joe12}.

Then it follows that the joint pmf is
\begin{equation}\label{jointserpmf}
f_{1\ldots d|t}(\y;\bbf,\gbf,\R)
=\int_{\mathcal{T}^{-1}(F_{1|t}^{-})}^{\mathcal{T}^{-1}(F_{1|t}^{+})}\cdots
\int_{\mathcal{T}^{-1}(F_{d|t}^{-})}^{\mathcal{T}^{-1}(F_{d|t}^{+})}  t_d(z_1,\ldots,z_d;\R) dz_1\ldots dz_d,\nonumber
\end{equation}
where $F_{j|t}^{-}:=F_{j|t}(y_{tj}-1|y_{t-1,j})$, $F_{1|t}^{+}:=F_{j|t}(y_{tj}|y_{t-1,j})$ and $t_d(\cdot;\R)$ denotes the MVT density  with latent correlation matrix $\R$ and $\nu$ degrees of freedom. The MVN case can be  treated as a special case of the MVT with a large value of $\nu$.

For the joint copula-based Markov model, we let $C_{j|t},\,j=1,\ldots,d$ and be parametric bivariate copulas, say with parameters $\theta_j,\,j=1,\ldots,d$ and $C$ be an MVT copula.  For the set of all parameters, let $\thbf=\{\bbf_j,\gbf_j,\theta_j,\R: j=1,\ldots,d\}$. We model the joint distribution in terms of $d$ bivariate copulas and an MVT copula.  Note that the copula $C_{j|t}$ models the time-series for the  $j$th response and the copula $C$ links that $j$ ordinal time-series responses.  Our general statistical model allows for selection of $C_{j|t}$  independently among a variety of parametric copula families, i.e., there are no constraints in the choices of parametric copulas $\{C_{j|t}: j=1,\ldots,d\}$.

\subsection{\label{concept}Conceptual framework}
\begin{figure}[!h]
\begin{center}
\begin{tabular}{cc}
\includegraphics[trim = 20mm 170mm 20mm 20mm, scale=.7]{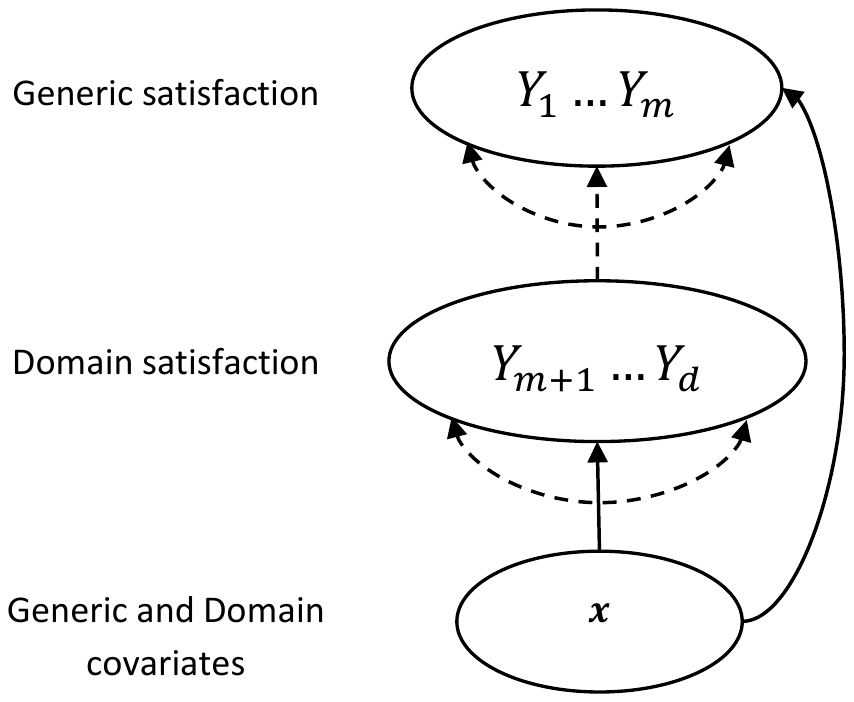}
\end{tabular}
\vspace{-1.5cm}
\caption{Life and domain satisfactions conceptual framework where solid and dotted lines indicate regression coefficients and latent correlations, respectively.} 
\label{fig:schematic}
\end{center}
\end{figure}

Let $(Y_1,\ldots,Y_m)$ and $(Y_{m+1},\ldots,Y_d)$ denote the generic and domain  satisfactions, respectively.  We propose an expansive set of relationships: objective individual characteristics and covariates  
directly relate to both generic and domain  satisfactions, with $(Y_{m+1},\ldots,Y_d)$ further influencing $(Y_1,\ldots,Y_m)$, while inter-dependencies  among them are also allowed. Such latent correlations capture the residual dependence (i.e. over and above the effect of covariates) among equations/outcomes. Figure \ref{fig:schematic} gives a schematic of our structural model. Such specification allows for direct, indirect and ripple (spill-over) effects on well-being, e.g. capturing at least three possible ways high individual income could affect generic satisfaction: a) a direct effect of income, b) an indirect effect through income satisfaction and c) an indirect effect through increased income satisfaction that itself is the result of an improved, for example, leisure satisfaction that was caused by the initial increase in income.

\section{\label{estimation}Estimation techniques and computational details}

For estimation purposes we propose a maximum simulated  likelihood  method, which is based  on evaluating the multidimensional integrals of the likelihood with randomized quasi Monte Carlo methods; an analysis of asymptotic properties of the estimators is shown in the  Appendix. 

\subsection{Simulated likelihood}

The log-likelihood of the joint copula-based Markov model is 
\begin{equation}\label{jointserlik}\ell_{1\ldots d|t}(\thbf)= \sum_{i=1}^n\left(\log f_{1\ldots d}(y_{i11},\ldots,y_{i1d};\thbf) + \sum_{t=2}^T\, \log f_{1\ldots d|t}(y_{it1},\ldots,y_{itd};\thbf)\right).
\end{equation}
where $f_{1\ldots d|t}(\cdot)$ is given in (\ref{jointserpmf}) and 
\begin{equation}\label{jointserpmf1}
f_{1\ldots d}(\y_1;\thbf)=
\int_{\mathcal{T}^{-1}[F_{1}(y_{1}-1;\bbf_1,\gbf)]}^{\mathcal{T}^{-1}[F_{1}(y_{1};\bbf_1,\gbf)]}\cdots
\int_{\mathcal{T}^{-1}[F_{d}(y_{d}-1;\bbf_d,\gbf)]}^{\mathcal{T}^{-1}[F_{d}(y_{d};\bbf_d,\gbf)]}  t_d(z_1,\ldots,z_d;\R) dz_1\ldots dz_d.
\end{equation}

We develop and implement a  maximum simulated likelihood estimator (MSLE). There exist general results on asymptotics of simulated likelihood based
estimators (see, e.g., \citealp{Gourieroux&Monfort-1991}). They usually involve a rate assumption on the number of simulations versus the sample size. 
Nevertheless, 
we propose  a simulated likelihood method, where the rectangle MVT probabilities in (\ref{jointserpmf}) and (\ref{jointserpmf1}) are computed using a quasi Monte Carlo method proposed by \cite{genz&bretz02}.
\cite{genz&bretz02} achieve error reduction of Monte Carlo methods through variance reduction techniques such as (a) transforming to a bounded integrand, (b) using antithetic variates, and (c) using a randomized quasi Monte Carlo method. The test results in \cite{genz&bretz02,genz&bretz2009} show that their method is very efficient, compared to other methods in the literature.
The method in \cite{genz&bretz02} is ``optimized" in the {\tt mtvnorm} R package \citep{genz-etal-2012}. Hence, on  the calculation of the MSLE,  one doesn't  need to worry about the selection, for example, of the number of simulated  quasi points.

 The estimated parameters can be obtained by  maximizing the simulated log-likelihood in (\ref{jointserlik}) over the model parameters $\thbf$. The method was initially proposed for the analysis of discrete (binary and count) longitudinal  data by \cite{nikoloulopoulos13b} and extended to a high-dimensional context in \cite{nikoloulopoulos2015a}.  We refer the interested reader to these papers for more details including studies of small-sample and asymptotic efficiency for Bernoulli, Poisson, and negative binomial regression models. In addition to that we study here the asymptotic properties of the maximum simulated likelihood estimators for ordinal regression models in an  Appendix.

\subsection{\label{sec-compdet}Computational details} 

The MSLEs can be derived using a three-step procedure:

\begin{enumerate}
\itemsep=0pt
\item For each $j$: 
\begin{enumerate}
\itemsep=0pt

\item Assuming time independence, $\ell_j(\bbf_j,\gbf_j)$ in (\ref{indlik}) is maximized over the univariate marginal parameters $\bbf_j,\gbf_j$.

\item Keeping the univariate  parameters $\bbf_j,\gbf_j$ fixed at the values estimated in (a), the $\ell_{j|t}(\bbf_j,\gbf_j,\th_j)$ in (\ref{serlik}) is maximized over the copula parameter $\th_j$.

\item Finally, using starting values from the estimates above the $\ell_{j|t}(\bbf_j,\gbf_j,\th_j)$ in (\ref{serlik}) is maximized over both the univariate $\bbf_j,\gbf_j$ and copula $\th_j$ parameters.

\end{enumerate}  

\item Setting all parameters to their estimated values from the first step, the $\ell_{1\ldots d|t}(\thbf)$ in (\ref{jointserlik}) is maximized over $\R$.

\item At the third and final step the  $\ell_{1\ldots d|t}(\thbf)$ in (\ref{jointserlik}) is maximized over $\thbf$ with initial parameters the estimates for the preceding steps.
\end{enumerate}

Given the typical large number of estimable  (univariate and copula) parameters  in multivariate models one can restrict themselves to only the first two steps of the method  to make inference computationally feasible. 
This two-step approach is  known in the copula literature as the Inference Function of Margins (IFM) method  \citep{joe&xu1996,joe97} and its asymptotic  
  efficiency  has been established \citep{joe05}. Hence using only the two first steps,  
\begin{itemize}
\item the model parameters can be  efficiently (in the sense of computing time and asymptotic variance) estimated; 
\item cross-model comparisons with respect to dependence structure and subsequently predictions and inferences can be performed. 
\end{itemize}  
 Note also in passing that compared to the (simulated) maximum likelihood, the IFM method  is not as punishing for misspecification of the dependence structure \citep{joe&xu1996,xu96}.

Each of the  estimated parameters can be obtained by using a quasi-Newton \citep{nash90} method applied to the log-likelihood. This numerical  method requires only the objective function, i.e., the joint log-likelihood, while the gradients
are computed numerically and the Hessian matrix of the second order derivatives is updated in each iteration.  
Since the estimation of parameters in MVT copula-based models is obtained using a quasi-Newton routine \citep{nash90}  
applied to the log-likelihood in (\ref{jointserlik}),  the use of randomized quasi Monte Carlo simulation to four decimal place
accuracy for evaluations of integrals works poorly, because numerical derivatives of the log-likelihood with respect to
the parameters are not smooth. In order to achieve smoothness, the same set of uniform random variables should be used for every rectangle probability that comes up in the optimization of the simulated likelihood   \citep{nikoloulopoulos13b,
nikoloulopoulos2015a}.

\section{\label{vuong-sec}Vuong's test for model comparison}

 A methodology for the comparison of non-nested models using the Vuong's test \citep{vuong1989} is formulated below to test if:

\begin{enumerate}
\item the copula-based Markov models with different choices of bivariate copulas outperform the copula-based Markov model with BVN (i.e. ``autoregressive-to-anything'' model);
\item the joint copula-based Markov models with MVT copulas provide better fit than their special case,  namely the MVN copula. 
\end{enumerate}

The Vuong's test is appropriate for parametric non-nested models comparisons and has often been used in the copula literature (e.g., \citealp{belgorodski10,Brechmann-Czado-Aas-2012,joe2014,Nikoloulopoulos2015b}). 

Assume models 1 and 2 with parametric densities $f^{(1)}$ and  $f^{(2)}$, respectively. Comparison of  
$$\Delta_{1f^\maltese}=N^{-1}\Bigl[\sum_i\{E_{f^\maltese}[\log f^\maltese(\cdot)]-E_{f^\maltese}[\log f^{(1)}(\cdot;\thbf^{(1)})]\}\Bigr],$$
and 
$$\Delta_{2f^\maltese}=N^{-1}\Bigl[\sum_i\{E_{f^\maltese}[\log f^\maltese(\cdot)]-E_{f^\maltese}[\log f^{(2)}(\cdot;\thbf^{(2)})]\}\Bigr],$$
where $\thbf^{(1)},\thbf^{(2)}$ are the parameters in models 1 and 2 respectively that lead to the closest Kullback-Leibler divergence to the true $f^\maltese$.  Model 1 is closer to the true $f^\maltese$, i.e., fits better if $\Delta=\Delta_{1f^\maltese}-\Delta_{2f^\maltese}<0$, while model 2 fits better if $\Delta>0$. The sample version of $\Delta$ with estimates $\hat\thbf^{(1)},\hat\thbf^{(2)}$ is
$$\bar D=\sum_{i=1}^N D_i/N,$$
where $D_i=\log\left[\frac{f^{(2)}\left(\cdot;\hat\thbf^{(2)}\right)}{f^{(1)}\left(\cdot;\hat\thbf^{(1)}\right)}\right]$. 

 \cite{vuong1989} 
has shown that asymptotically under the null hypothesis $H_0:\Delta=0$, i.e.,  models 1 and 2 have the same  parametric densities $f^{(1)}$ and  $f^{(2)}$,  
$$z_0=\sqrt{N}\bar D/s\widesim{H_0}\mathcal{N}(0,1),$$
where  $s^2=\frac{1}{N-1}\sum_{i=1}^N(D_i-\bar D)^2$. 
Rejection on the null hypothesis follows if $\abs{z_0}$ is greater than the critical value from the standard normal distribution, denoted $\mathcal{N}(0,1)$.

\section{\label{sec-appl} British Household Panel Survey}

For the estimation of the model in Figure \ref{fig:schematic} we use data from the British Household Panel Survey (BHPS). The BHPS was an annual longitudinal survey (now superseded by Understanding Society) carried out by the Institute for Social and Economic Research sampling about 10,000 individuals aged 16 years or over from 1991 to 2008. However, the survey modules required for our model are collected on a bi-annual basis starting in wave 6 (i.e. 1996) and finishing in wave 18 (i.e. 2008), resulting in a maximum of 7 measurements per individual.     
To ease coding, we use individuals observed at all seven time points resulting in a sample size of 4186 individuals, though our methodology does not depend on a constant ``cluster" size $T$.

\subsection{Equations and covariates}

The actual model estimated is based on one generic $Y_1$ and six domain satisfaction $(Y_2,Y_3,\ldots,Y_7)$ questions each answered on a 1 (not satisfied at all) to 6 (completely satisfied) likert scale. The seven equations (outcomes) are:

\begin{description}
\itemsep=0pt
	\item[$Y_1$:] Satisfaction with Life overall  
	\item[$Y_2$:] Satisfaction with Health  
	\item[$Y_3$:] Satisfaction with Income  
	\item[$Y_4$:] Satisfaction with House/flat  
	\item[$Y_5$:] Satisfaction with Spouse/partner  
	\item[$Y_6$:] Satisfaction with Job  
	\item[$Y_7$:] Satisfaction with Leisure 
\end{description}

Each of $(Y_2,Y_3,\ldots,Y_7)$ is conditioned upon  individual characteristics  (i.e. age, age square, gender, household size, number of kids, education and geographical region within the UK) that are common controls in the well-being literature and domain specific factors that appear only on the respective domain equation  and would enhance the ability to capture domain specific variation. 
For example, number of health problems are used for the satisfaction with health equation, disaggregated sources of income in the satisfaction with income equation and so on for the rest of the equations. Table \ref{tab:desstat} offers definitions and breakdowns for domain specific variables, as well as the reference levels for all categorical variables. Note that the original {\emph{region}} variable is a geographical identifier that splits UK into 19 areas and which is aggregated into 5 ``super''-regions for the regressions. Further, income rank captures relative income and is determined by the rank of the individual in their original region (i.e. one of 19 areas) according to their total annual income.

For equation $Y_1$ we use as covariates all common and domain specific variables capturing the direct effects that each characteristic has on overall life satisfaction. This formulation allows the separate identification of the direct effect of 
 domain characteristics and the effect of composite domains on generic satisfaction, while, as the econometric model poses, residual dependence between and among $Y_1$ and $(Y_2,Y_3,\ldots,Y_7)$ is captured by estimable latent correlation parameters. 

Given the multivariate nature of our model we restrict the estimation sample to those for whom information on each equation is available, i.e. married, employed individuals (up to 70 years old) who indicate to have a partner. Further, as discussed earlier, albeit not a requirement of the model, we only keep those individuals that appear in all seven waves of  the data. Descriptive statistics for all equation outcomes, and covariates are available in Table \ref{tab:desstat} where informal sample selection comparisons can also be drawn between the estimation and full sample.   
Respondents tend to be most satisfied with their partners, followed by satisfaction with their houses and then with life overall. Income and leisure are the least satisfactory dimensions. Overall, our sample is 46\% males and on average 45 years old, with 12\% having a higher  degree and 7\% residing in London, 20\% in South England, 16\% in the Midlands and 19\% in North England. Comparing restricted and full samples the former is slightly less educated and with a different income sources structure but on average individual characteristics and domain covariates are largely comparable across the two samples. 

\begin{table}[!b]
\centering
\caption{\label{tab:desstat}Descriptive statistics for generic and domain satisfaction outcomes and all common and domain specific covariates for estimation and full samples.}
\begin{tabular}{lccc}
\hline & Estimation sample &  & Full sample \\ \cline{2-2} \cline{4-4} \\ 

Equations &  &  &  \\ \cline{1-1} 
$Y_1$: Satisfaction with Life overall  & 4.25 &  & 4.29 \\
$Y_2$: Satisfaction with Health  & 4.00 &  & 4.23 \\
$Y_3$: Satisfaction with Income  & 3.64 &  & 3.78 \\
$Y_4$: Satisfaction with House/flat  & 4.45 &  & 4.44 \\
$Y_5$: Satisfaction with Spouse/partner  & 5.24 &  & 5.27 \\
$Y_6$: Satisfaction with Job  & 4.05 &  & 4.01 \\
$Y_7$: Satisfaction with Leisure  & 3.93 &  & 3.77 \\
 & & & \\
Covariates &  &  &  \\ \cline{1-1} 
Age & 4.53 &  & 4.36 \\
Age$^2$ & 23.98 &  & 19.72 \\
Sex (1 if male) & 0.46 &  & 0.55 \\
Household size  & 0.29 &  & 0.33 \\
\# of kids & 0.06 &  & 0.09 \\
Education (ref category: Uni degree) &  &  &  \\
\hspace{2ex}hnd,hnc, a/o levels, cse & 0.55 &  & 0.69 \\
\hspace{2ex}No education & 0.33 &  & 0.11 \\
Region (ref category: London) &  &  &  \\
\hspace{2ex}South  & 0.20 &  & 0.30 \\
\hspace{2ex}Midlands  & 0.16 &  & 0.22 \\
\hspace{2ex}North  & 0.19 &  & 0.32 \\
\hspace{2ex}RUK  & 0.38 &  & 0.10 \\
Health &  &  &  \\
\hspace{2ex}\# Health problems  & 0.12 &  & 0.08 \\
Income &  &  &  \\
\hspace{2ex}ln(Labour Income) & 7.53 &  & 10.47 \\
\hspace{2ex}ln(Pension Income) & 1.98 &  & 0.68 \\
\hspace{2ex}ln(Benefit Income) & 5.89 &  & 4.80 \\
\hspace{2ex}ln(Transfer Income) & 0.70 &  & 0.41 \\
\hspace{2ex}ln(Investment Income) & 3.46 &  & 4.19 \\
\hspace{2ex}Regional incomne rank (stadardized) & 0.00 &  & -0.07 \\
House type (ref category: Detached) &  &  &  \\
\hspace{2ex}Semi-detached  & 0.32 &  & 0.43 \\
\hspace{2ex}Terraced  & 0.27 &  & 0.21 \\
\hspace{2ex}Other  & 0.15 &  & 0.02 \\
House value (ref category: 0-50K) &  &  &  \\
\hspace{2ex}50K-100K  & 0.35 &  & 0.28 \\
\hspace{2ex}100K-175K & 0.23 &  & 0.25 \\
\hspace{2ex}175K-250K & 0.14 &  & 0.19 \\
\hspace{2ex}$>$250K & 0.11 &  & 0.15 \\

\hline \multicolumn{4}{r}{{Continued on next page}} 
\end{tabular}
\end{table}

\setcounter{table}{0}

\begin{table}[!t]
\centering
\caption{Continued.}
\begin{tabular}{lccc}
\hline & Estimation sample &  & Full sample \\ \cline{2-2} \cline{4-4} \\ 

Spouse Charateristics &  &  &  \\
\hspace{2ex} Age of spouse & 4.70 &  & 4.37 \\
\hspace{2ex} Age$^2$ of spouse  & 24.46 &  & 19.92 \\
\hspace{2ex} Sex of spouse  & 0.50 &  & 0.45 \\
\hspace{2ex} Education (ref category: Uni degree) &  &  &  \\
\hspace{4ex} hnd,hnc, a/o levels, cse & 0.55 &  & 0.67 \\
\hspace{4ex} No education & 0.32 &  & 0.16 \\
\hspace{2ex}ln(Labour Income) & 6.35 &  & 8.46 \\
\hspace{2ex}\# Health problems & 0.11 &  & 0.09 \\
\hspace{2ex}Hours of wk housework & 1.19 &  & 1.18 \\
Satisfaction with job pay (ref category: Low) &  &  &  \\
\hspace{2ex}Medium & 0.33 &  & 0.33 \\
\hspace{2ex}Very  & 0.45 &  & 0.49 \\
Satisfaction with job security (ref category: Low) &  &  &  \\
\hspace{2ex}Medium & 0.26 &  & 0.31 \\
\hspace{2ex}Very  & 0.61 &  & 0.57 \\
Satisfaction with work itself (ref category: Low) &  &  &  \\
\hspace{2ex}Medium & 0.28 &  & 0.30 \\
\hspace{2ex}Very  & 0.62 &  & 0.61 \\
Satisfaction with hours worked (ref category: Low) &  &  &  \\
\hspace{2ex}Medium & 0.31 &  & 0.33 \\
\hspace{2ex}Very  & 0.54 &  & 0.51 \\
Leisure activities (1 if several times a year) &  &  &  \\
\hspace{2ex}walk/swim/play sport  & 0.67 &  & 0.75 \\
\hspace{2ex}watch live sport  & 0.25 &  & 0.33 \\
\hspace{2ex}cinema & 0.47 &  & 0.53 \\
\hspace{2ex}theatre/concert  & 0.36 &  & 0.37 \\
\hspace{2ex}out for a drink & 0.69 &  & 0.84 \\
\hspace{2ex}work in garden & 0.64 &  & 0.86 \\
\hspace{2ex}diy, car maintenance & 0.53 &  & 0.72 \\
\hspace{2ex}attend evening classes  & 0.27 &  & 0.30 \\
\hspace{2ex}attend local groups & 0.20 &  & 0.21 \\ \hline 
\end{tabular}
\end{table}

\subsection{Fitted copula-based Markov models for each ordinal time-series}

We fit the 
copula-based Markov  model with BVN, Gumbel, s.Gumbel, and BVT bivariate linking copulas. For BVT, choices of $\nu$ were $1,2,\ldots,10$.  To make it easier to compare the dependence parameters, we convert the estimated parameters to Kendall's $\tau$'s in $(0,1)$  via the relations 
$\tau=\frac{2}{\pi}\arcsin{\theta}$,
$\tau=1+4\theta^{-1}\left[\frac{1}{\theta}\int_0^{\theta}\frac{t}{e^t-1}dt-1\right],$
and 
$\tau=1-\th^{-1}$
for elliptical, Frank and Gumbel copulas in \cite{HultLindskog02}, \cite{genest87}, and \cite{genest&mackay86}, respectively. 
Note that Kendall's tau only accounts for the dependence dominated by the middle of the data, and it is expected to be similar amongst  different families of copulas. However, the tail dependence varies, as explained in Section \ref{sec-families}, and is a property to consider when choosing amongst different families of copulas. For the  model with BVT we summarize the choice of integer $\nu$ with the largest log-likelihood.

Given  the equality in  number of parameters between models, the log-likelihood at estimates can be used as a measure for goodness of fit across all models. We further compute the \nocite{vuong1989}Vuong's  (1989) test to formally assess if more probability is accumulated in the joint tails than one would expect via  a BVN  copula.

For these data, if a respondent  thinks about the maximum or minimum satisfaction at year $t$ it seems natural to think about the maximum or minimum satisfaction at year $t+1$ and year $t-1$.  That is, based on data descriptions, we could expect a priori that a  model with $C_{j|t}$ being the BVT copulas might be plausible, as data have more probability in the joint tails.

\begin{table}[!b]
\centering \caption{\label{tab:lfsato} Estimated parameters and joint log-likelihoods $\ell_{j|t}$ for the copula-based Markov models for ordinal time-series with covariates for  satisfaction with life overall $Y_1$, where a parametric copula family $C_{j|t}$ is used for the joint distribution of subsequent observations. For the best fit, the standard errors (SE) of the estimates, Wald tests ($Z$) and $p$-values are also presented. }

\begin{scriptsize}
\begin{tabular}{lcccccccc} \hline 
 & BVN  &  Frank  &  Gumbel  &  s.Gumbel  &  \multicolumn{4}{c}{BVT, $\nu=4$} \\ \cline{2-9}
 & Est.  &  Est.  &  Est.  &  Est.  &  Est.  & SE  & $Z$  & $p$-value  \\ \cline{2-9}
$\a_{1}$ & -2.93 & -3.24 & -2.88 & -2.63 & -2.63 & 0.83 & -3.18 & 0.01 \\
$\a_{2}$ & -2.07 & -2.38 & -1.98 & -1.90 & -1.81 & 0.82 & -2.20 & 0.03 \\
$\a_{3}$ & -1.33 & -1.64 & -1.22 & -1.23 & -1.09 & 0.82 & -1.32 & 0.19 \\
$\a_{4}$ & -0.14 & -0.47 & -0.03 & -0.09 & 0.09 & 0.83 & 0.11 & 0.92 \\
$\a_{5}$ & 1.44 & 1.10 & 1.48 & 1.50 & 1.66 & 0.83 & 2.01 & 0.04 \\
Age & -0.74 & -0.84 & -0.58 & -0.70 & -0.55 & 0.31 & -1.79 & 0.07 \\
Age$^2$ & 0.10 & 0.11 & 0.08 & 0.10 & 0.08 & 0.03 & 2.23 & 0.03 \\
Sex & 0.13 & 0.45 & 0.32 & -0.09 & 0.08 & 0.37 & 0.20 & 0.84 \\
Household size & -0.72 & -0.70 & -0.78 & -0.64 & -0.70 & 0.32 & -2.16 & 0.03 \\
\# of kids & 0.10 & 0.01 & -0.03 & 0.08 & -0.03 & 0.34 & -0.10 & 0.93 \\
Education &  &  &  &  &  &  &  &   \\
\hspace{2ex}No education & 0.08 & 0.09 & 0.09 & 0.11 & 0.11 & 0.11 & 1.06 & 0.29 \\
\hspace{2ex} hnd,hnc, a/o levels, cse & 0.03 & 0.03 & 0.05 & 0.03 & 0.05 & 0.07 & 0.73 & 0.46 \\
Region &  &  &  &  &  &  &  &   \\
\hspace{2ex}South  & 0.22 & 0.25 & 0.28 & 0.18 & 0.24 & 0.11 & 2.08 & 0.04 \\
\hspace{2ex}Midlands  & 0.12 & 0.18 & 0.17 & 0.09 & 0.12 & 0.12 & 1.03 & 0.30 \\
\hspace{2ex}North   & 0.16 & 0.15 & 0.22 & 0.16 & 0.21 & 0.12 & 1.82 & 0.07 \\
\hspace{2ex}RUK & 0.30 & 0.32 & 0.29 & 0.31 & 0.29 & 0.16 & 1.85 & 0.07 \\
Health &  &  &  &  &  &  &  &    \\
\hspace{2ex}\# Health problems  & -1.43 & -1.53 & -1.46 & -1.24 & -1.34 & 0.19 & -7.05 & 0.00 \\
Income &  &  &  &  &  &  &  &   \\
\hspace{2ex}ln(Labour Income) & 0.05 & 0.01 & 0.03 & 0.07 & 0.05 & 0.06 & 0.95 & 0.34 \\
\hspace{2ex}ln(Pension Income) & 0.01 & 0.00 & 0.00 & 0.00 & 0.00 & 0.01 & 0.20 & 0.84 \\
\hspace{2ex}ln(Benefit Income) & -0.01 & -0.00 & -0.01 & -0.01 & -0.01 & 0.01 & -1.01 & 0.31 \\
\hspace{2ex}ln(Transfer Income) & -0.01 & -0.01 & -0.01 & -0.00 & -0.00 & 0.01 & -0.29 & 0.77 \\
\hspace{2ex}ln(Investment Income) & 0.01 & 0.01 & 0.01 & 0.01 & 0.01 & 0.01 & 1.29 & 0.20 \\
\hspace{2ex}Regional income rank (standardized) & -0.01 & 0.02 & 0.01 & -0.03 & -0.02 & 0.05 & -0.33 & 0.74 \\
House type &  &  &  &  &  &  &  &   \\
\hspace{2ex}Semi-detached & -0.03 & -0.049 & -0.04 & -0.04 & -0.04 & 0.05 & -0.80 & 0.42 \\
\hspace{2ex}Terraced & -0.09 & -0.12 & -0.09 & -0.11 & -0.10 & 0.07 & -1.56 & 0.12 \\
\hspace{2ex}Other & -0.45 & -0.44 & -0.41 & -0.44 & -0.43 & 0.15 & -2.80 & 0.01 \\
House value &  &  &  &  &  &  &  &    \\
\hspace{2ex}50K-100K & -0.06 & -0.03 & -0.06 & -0.07 & -0.07 & 0.06 & -1.21 & 0.23 \\
\hspace{2ex}100K-175K & -0.12 & -0.07 & -0.10 & -0.15 & -0.13 & 0.07 & -1.79 & 0.07 \\
\hspace{2ex}175K-250K & -0.15 & -0.12 & -0.12 & -0.17 & -0.15 & 0.08 & -1.82 & 0.07 \\
\hspace{2ex}$>$250K & -0.12 & -0.09 & -0.09 & -0.14 & -0.12 & 0.10 & -1.20 & 0.23 \\
Spouse Charateristics &  &  &  &  &  &  &  &    \\
\hspace{2ex} Age & 0.07 & -0.03 & 0.03 & 0.07 & 0.04 & 0.28 & 0.13 & 0.90 \\
\hspace{2ex} Age$^2$ & -0.03 & -0.01 & -0.02 & -0.03 & -0.03 & 0.03 & -0.81 & 0.42 \\
\hspace{2ex} Sex & 0.13 & 0.46 & 0.31 & -0.07 & 0.08 & 0.37 & 0.22 & 0.83 \\
\hspace{2ex}Education &  &  &  &  &  &  &  &   \\
\hspace{4ex}No education & 0.10 & 0.06 & 0.10 & 0.12 & 0.12 & 0.10 & 1.16 & 0.25 \\
\hspace{4ex}hnd,hnc, a/o levels, cse & -0.05 & -0.03 & -0.09 & -0.04 & -0.09 & 0.07 & -1.15 & 0.25 \\
\hspace{2ex}ln(Labour Income) of spouse & 0.01 & 0.01 & 0.01 & 0.01 & 0.01 & 0.01 & 1.17 & 0.24 \\
\hspace{2ex}\# Health problems of spouse & -0.27 & -0.25 & -0.19 & -0.26 & -0.19 & 0.17 & -1.11 & 0.27 \\
\hspace{2ex}Hours of wk housework & -0.02 & -0.02 & -0.02 & -0.01 & -0.01 & 0.02 & -0.54 & 0.59 \\
Satisfaction with job pay &  &  &  &  &  &  &  &    \\
\hspace{2ex}Medium  & 0.14 & 0.18 & 0.15 & 0.13 & 0.14 & 0.04 & 3.15 & 0.00 \\
\hspace{2ex}Very & 0.23 & 0.28 & 0.22 & 0.20 & 0.20 & 0.05 & 4.44 & 0.00 \\
Satisfaction with job security &  &  &  &  &  &  &  &    \\
\hspace{2ex}Medium & 0.10 & 0.10 & 0.09 & 0.11 & 0.10 & 0.05 & 2.14 & 0.03 \\
\hspace{2ex}Very & 0.18 & 0.18 & 0.16 & 0.17 & 0.16 & 0.05 & 3.42 & 0.00 \\
Satisfaction with work itself &  &  &  &  &  &  &  &    \\
\hspace{2ex}Medium & 0.14 & 0.16 & 0.12 & 0.14 & 0.14 & 0.05 & 2.62 & 0.01 \\
\hspace{2ex}Very & 0.41 & 0.41 & 0.37 & 0.39 & 0.38 & 0.05 & 7.03 & 0.00 \\\hline
 \multicolumn{9}{r}{{Continued on next page}}
\end{tabular}
\end{scriptsize}
\end{table}

\setcounter{table}{1}

\begin{table}[!t]
\centering
\caption{Continued.}
\begin{scriptsize}
\begin{tabular}{lcccccccc} \hline 
& BVN  &  Frank  &  Gumbel  &  s.Gumbel  &  \multicolumn{4}{c}{BVT, $\nu=4$} \\ \cline{2-9}
 & Est.  &  Est.  &  Est.  &  Est.  &  Est.  & SE  & $Z$  & $p$-value  \\ \cline{2-9}

Satisfaction with hours worked &  &  &  &  &  &  &  &   \\
 \hspace{2ex}Medium & 0.21 & 0.19 & 0.19 & 0.20 & 0.19 & 0.04 & 4.32 & 0.00 \\
\hspace{2ex}Very & 0.32 & 0.29 & 0.30 & 0.29 & 0.30 & 0.05 & 6.21 & 0.00 \\
Frequency of leisure activities &  &  &  &  &  &  &  &   \\
\hspace{2ex}walk/swim/play sport & -0.02 & 0.00 & -0.02 & -0.03 & -0.04 & 0.04& -1.07 & 0.29 \\
\hspace{2ex}watch live sport & 0.06& 0.06 & 0.05 & 0.06 & 0.05 & 0.04 & 1.18 & 0.24 \\
\hspace{2ex}cinema & -0.03 & -0.03 & -0.02 & -0.04 & -0.03 & 0.04 & -0.99 & 0.32 \\
\hspace{2ex}theatre/concert & 0.01 & -0.01 & -0.00 & 0.02 & 0.01& 0.03 & 0.25 & 0.80\\
\hspace{2ex}out for a drink  & 0.06 & 0.06 & 0.07 & 0.05 & 0.06 & 0.05 & 1.25 & 0.21 \\
\hspace{2ex}work in garden  & 0.06 & 0.05 & 0.04 & 0.04 & 0.03 & 0.05 & 0.56 & 0.58 \\
\hspace{2ex}diy, car maintenance & 0.01 & 0.01 & 0.02 & -0.01 & 0.00 & 0.04 & -0.01 & 0.10 \\
\hspace{2ex}attend evening classes & 0.04 & 0.05 & 0.05 & 0.04 & 0.05 & 0.03 & 1.50 & 0.13 \\
\hspace{2ex}attend local groups & 0.01 & 0.00 & 0.01 & 0.01 & 0.01 & 0.04 & 0.22 & 0.82 \\ \hline
$\tau$ & 0.36 & 0.40 & 0.39 & 0.37 & 0.38 & 0.01 & 28.5 & 0.00 \\
$-\ell_{j|t}$ & 4734.1 & 4734.2 & 4719.1 & 4740.0 & \multicolumn{4}{c}{4689.2}  \\ \hline
$z_0$ ($p$-value)&{-}& -0.010 (0.992)	&1.159	(0.246)	&-0.576	(0.565)	&\multicolumn{4}{c}{3.419 (0.001)}\\\hline
\end{tabular}
\end{scriptsize}
\end{table}

%%%%%
\begin{table}[!h]
\centering \caption{\label{tab:lfsat1} Estimated parameters and joint log-likelihoods $\ell_{j|t}$ for the copula-based Markov models for ordinal time-series with covariates for satisfaction with health $Y_2$, where a parametric copula family $C_{j|t}$ is used for the joint distribution of subsequent observations. For the best fit, the standard errors (SE) of the estimates, Wald tests ($Z$) and $p$-values are also presented.}

\begin{scriptsize}
\begin{tabular}{lcccccccc} \hline 
 & BVN & Frank & Gumbel & s.Gumbel & \multicolumn{4}{c}{BVT, $\nu=6$} \\ \cline{2-9}
  & Est.  &  Est.  &  Est.  &  Est.  &  Est.  & SE  & $Z$  & $p$-value  \\ \cline{2-9}
$\a_1$ & -3.85 & -3.85 & -4.01 & -3.38 & -3.75 & 0.46 & -8.20 & 0.00 \\
$\a_2$ & -3.21 & -3.20 & -3.35 & -2.82 & -3.12 & 0.46 & -6.87 & 0.00 \\
$\a_3$ & -2.67 & -2.65 & -2.79 & -2.32 & -2.59 & 0.45 & -5.69 & 0.00 \\
$\a_4$ & -1.86 & -1.85 & -1.98 & -1.54 & -1.78 & 0.45 & -3.92 & 0.00 \\
$\a_5$ & -0.64 & -0.65 & -0.81 & -0.31 & -0.56 & 0.45 & -1.24 & 0.22 \\
Age & -0.75 & -0.72 & -0.82 & -0.63 & -0.74 & 0.21 & -3.49 & 0.00 \\
Age$^2$ & 0.08 & 0.08 & 0.09 & 0.07 & 0.08 & 0.02 & 3.44 & 0.00 \\
Sex & 0.02 & 0.01 & 0.01 & 0.01 & 0.02 & 0.05 & 0.32 & 0.75 \\
Household size & -0.49 & -0.44 & -0.31 & -0.51 & -0.42 & 0.29 & -1.47 & 0.14 \\
\# of kids & 0.36 & 0.26 & 0.22 & 0.43 & 0.34 & 0.31 & 1.10 & 0.27 \\
Education &  &  &  &  &  &  &  &  \\
\hspace{2ex}No education & 0.00 & -0.04 & 0.01 & 0.03 & 0.03 & 0.09 & 0.33 & 0.74 \\
\hspace{2ex} hnd,hnc, a/o levels, cse & 0.02 & 0.02 & 0.02 & 0.02 & 0.02 & 0.06 & 0.38 & 0.70 \\
Region &  &  &  &  &  &  &  &  \\
\hspace{2ex}South & 0.19 & 0.12 & 0.17 & 0.20 & 0.20 & 0.10 & 1.97 & 0.05 \\
\hspace{2ex}Midlands & 0.10 & 0.05 & 0.09 & 0.11 & 0.12 & 0.11 & 1.08 & 0.28 \\
\hspace{2ex}North  & 0.07 & 0.01 & 0.04 & 0.11 & 0.09 & 0.10 & 0.87 & 0.38 \\
\hspace{2ex}RUK & 0.24 & 0.24 & 0.20 & 0.25 & 0.23 & 0.12 & 1.93 & 0.05 \\
Health &  &  &  &  &  &  &  &  \\
\hspace{2ex}\# Health problems & -3.59 & -3.45 & -3.34 & -3.45 & -3.42 & 0.20 & -16.8 & 0.00 \\ \hline
$\tau$  & 0.34 & 0.37 & 0.36 & 0.34 & 0.35 & 0.01 & 28.3 & 0.00 \\
$-\ell_{j|t}$ & 5752.2 & 5741.8 & 5739.9 & 5773.2 & \multicolumn{4}{c}{5725.9} \\ \hline
$z_0$ ($p$-value)&{-}&1.112	(0.266)&	1.147	(0.251)&	-1.916	(0.055)&	\multicolumn{4}{c}{3.009 	(0.003)}\\\hline
\end{tabular}
\end{scriptsize}
\end{table}

%%%%%
\begin{table}[!h]
\centering \caption{\label{tab:lfsat2} Estimated parameters and joint log-likelihoods $\ell_{j|t}$ for the copula-based Markov models for ordinal time-series with covariates for   satisfaction with income $Y_3$, where a parametric copula family $C_{j|t}$ is used for the joint distribution of subsequent observations. For the best fit, the standard errors (SE) of the estimates, Wald tests ($Z$) and $p$-values are also presented.}

\begin{scriptsize}
\begin{tabular}{lcccccccc} \hline 
& BVN & Frank & Gumbel & s.Gumbel & \multicolumn{4}{c}{BVT, $\nu=4$} \\ \cline{2-9}
& Est.  &  Est.  &  Est.  &  Est.  &  Est.  & SE  & $Z$  & $p$-value  \\ \cline{2-9} 
$\a_1$ & 1.79 & 1.26 & 1.84 & 1.57 & 1.83 & 0.67 & 2.76 & 0.01 \\
$\a_2$ & 2.42 & 1.89 & 2.49 & 2.13 & 2.45 & 0.67 & 3.67 & 0.00 \\
$\a_3$ & 3.06 & 2.51 & 3.15 & 2.73 & 3.09 & 0.67 & 4.62 & 0.00 \\
$\a_4$ & 4.04 & 3.46 & 4.11 & 3.71 & 4.07 & 0.67 & 6.07 & 0.00 \\
$\a_5$ & 5.18 & 4.60 & 5.14 & 4.90 & 5.20 & 0.67 & 7.71 & 0.00 \\
Age & -0.39 & -0.52 & -0.35 & -0.24 & -0.23 & 0.22 & -1.06 & 0.29 \\
Age$^2$ & 0.04 & 0.06 & 0.03 & 0.03 & 0.02 & 0.02 & 0.86 & 0.39 \\
Sex & -0.01 & -0.02 & -0.01 & -0.03 & -0.02 & 0.05 & -0.45 & 0.66 \\
Household size & -1.44 & -1.30 & -1.36 & -1.23 & -1.27 & 0.30 & -4.19 & 0.00 \\
\# of kids & 0.85 & 0.89 & 0.73 & 0.70 & 0.68 & 0.32 & 2.13 & 0.03 \\
Education &  &  &  &  &  &  &  &  \\
\hspace{2ex}No education & -0.19 & -0.17 & -0.12 & -0.24 & -0.20 & 0.10 & -1.95 & 0.05 \\
\hspace{2ex} hnd,hnc, a/o levels, cse & -0.16 & -0.15 & -0.11 & -0.18 & -0.15 & 0.07 & -2.22 & 0.03 \\
Region &  &  &  &  &  &  &  &  \\
\hspace{2ex}South & 0.08 & 0.06 & 0.18 & -0.04 & 0.07 & 0.12 & 0.56 & 0.57 \\
\hspace{2ex}Midlands & 0.04 & -0.04 & 0.17 & -0.09 & 0.05 & 0.12 & 0.46 & 0.65 \\
\hspace{2ex}North  & 0.00 & -0.07 & 0.14 & -0.15 & 0.01 & 0.12 & 0.05 & 0.96 \\
\hspace{2ex}RUK & -0.02 & -0.13 & 0.12 & -0.14 & 0.01 & 0.15 & 0.08 & 0.94 \\
Income &  &  &  &  &  &  &  &  \\
\hspace{2ex}ln(Labour Income)  & 0.46 & 0.43 & 0.45 & 0.40 & 0.43 & 0.05 & 8.73 & 0.00 \\
\hspace{2ex}ln(Pension Income) & 0.02 & 0.02 & 0.02 & 0.02 & 0.02 & 0.01 & 2.31 & 0.02 \\
\hspace{2ex}ln(Benefit Income) & -0.01 & -0.01 & -0.01 & -0.01 & -0.01 & 0.01 & -2.11 & 0.03 \\
\hspace{2ex}ln(Transfer Income) & -0.01 & -0.01 & -0.01 & -0.01 & -0.01 & 0.01 & -1.07 & 0.29 \\
\hspace{2ex}ln(Investment Income) & 0.04 & 0.03 & 0.03 & 0.03 & 0.03 & 0.01 & 5.84 & 0.00 \\
\hspace{2ex}Regional income rank (standardized) & 0.02 & 0.02 & 0.00 & 0.02 & 0.01 & 0.04 & 0.13 & 0.89 \\ \hline
$\tau$  & 0.40 & 0.43 & 0.42 & 0.41 & 0.41 & 0.01 & 33.2 & 0.00 \\
$-\ell_{j|t}$ & 5750.1 & 5756.6 & 5756.3 & 5726.6 & \multicolumn{4}{c}{5698.1} \\ \hline
$z_0$ ($p$-value)&{-}&-0.549	(0.583)&	-0.526	(0.599)	&2.008	(0.045)&	\multicolumn{4}{c}{4.272 	(0.000)}\\\hline
\end{tabular}
\end{scriptsize}
\end{table}

%%%%%
\begin{table}[!h]
\centering \caption{\label{tab:lfsat3} Estimated parameters and joint log-likelihoods $\ell_{j|t}$ for the copula-based Markov models for ordinal time-series with covariates for  satisfaction with house/flat $Y_4$, where a parametric copula family $C_{j|t}$ is used for the joint distribution of subsequent observations. For the best fit, the standard errors (SE) of the estimates, Wald tests ($Z$) and $p$-values are also presented.}

\begin{scriptsize}
\begin{tabular}{lcccccccc} \hline
 & BVN & Frank & Gumbel & s.Gumbel & \multicolumn{4}{c}{BVT, $\nu=5$} \\ \cline{2-9}
   & Est.  &  Est.  &  Est.  &  Est.  &  Est.  & SE  & Z  & $p$-value  \\ \cline{2-9}
$\a_1$ & -3.19 & -3.36 & -3.22 & -2.71 & -2.97 & 0.47 & -6.36 & 0.00 \\
$\a_2$ & -2.65 & -2.79 & -2.65 & -2.27 & -2.46 & 0.47 & -5.27 & 0.00 \\
$\a_3$ & -2.11 & -2.25 & -2.09 & -1.80 & -1.94 & 0.47 & -4.15 & 0.00 \\
$\a_4$ & -1.21 & -1.36 & -1.17 & -0.94 & -1.04 & 0.47 & -2.22 & 0.03 \\
$\a_5$ & 0.06 & -0.10 & 0.05 & 0.35 & 0.24 & 0.47 & 0.51 & 0.61 \\
Age & -0.48 & -0.54 & -0.41 & -0.39 & -0.39 & 0.22 & -1.78 & 0.07 \\
Age$^2$ & 0.06 & 0.07 & 0.05 & 0.06 & 0.05 & 0.02 & 2.11 & 0.04 \\
Sex & 0.02 & 0.01 & 0.00 & 0.03 & 0.01 & 0.05 & 0.24 & 0.81 \\
Household size & -0.70 & -0.72 & -0.75 & -0.47 & -0.57 & 0.29 & -1.95 & 0.05 \\
\# of kids & 0.27 & 0.39 & 0.32 & 0.10 & 0.17 & 0.32 & 0.55 & 0.58 \\
Education &  &  &  &  &  &  &  &  \\
\hspace{2ex}No education & 0.36 & 0.33 & 0.37 & 0.26 & 0.32 & 0.10 & 3.25 & 0.00 \\
\hspace{2ex} hnd,hnc, a/o levels, cse & 0.03 & 0.04 & 0.01 & 0.03 & 0.01 & 0.06 & 0.16 & 0.87 \\
Region &  &  &  &  &  &  &  &  \\
\hspace{2ex}South & -0.03 & 0.00 & -0.01 & -0.07 & -0.06 & 0.11 & -0.53 & 0.60 \\
\hspace{2ex}Midlands & -0.05 & 0.01 & 0.00 & -0.11 & -0.09 & 0.12 & -0.73 & 0.46 \\
\hspace{2ex}North  & -0.03 & 0.03 & -0.01 & -0.12 & -0.09 & 0.11 & -0.82 & 0.41 \\
\hspace{2ex}RUK & 0.07 & 0.15 & 0.09 & 0.00 & 0.01 & 0.13 & 0.10 & 0.92 \\
House Type &  &  &  &  &  &  &  &  \\
\hspace{2ex}Semi-detached & -0.27 & -0.29 & -0.29 & -0.21 & -0.25 & 0.05 & -4.75 & 0.00 \\
\hspace{2ex}Terraced & -0.49 & -0.50 & -0.50 & -0.42 & -0.47 & 0.07 & -7.21 & 0.00 \\
\hspace{2ex}Other  & -1.15 & -1.26 & -1.07 & -1.05 & -1.06 & 0.15 & -7.10 & 0.00 \\
House value &  &  &  &  &  &  &  &  \\
\hspace{2ex}50K-100K & 0.16 & 0.12 & 0.13 & 0.15 & 0.14 & 0.06 & 2.34 & 0.02 \\
\hspace{2ex}100K-175K & 0.25 & 0.18 & 0.21 & 0.20 & 0.19 & 0.07 & 2.87 & 0.00 \\
\hspace{2ex}175K-250K & 0.31 & 0.24 & 0.26 & 0.25 & 0.25 & 0.08 & 3.24 & 0.00 \\
\hspace{2ex}$>$250K & 0.42 & 0.40 & 0.37 & 0.32 & 0.32 & 0.09 & 3.65 & 0.00 \\ \hline
$\tau$ & 0.35 & 0.38 & 0.37 & 0.36 & 0.37 & 0.01 & 29.28 & 0.00 \\
$-\ell_{j|t}$ & 5407.6 & 5394.8 & 5396.3 & 5411.4 & \multicolumn{4}{c}{5373.0} \\ \hline
$z_0$ ($p$-value) &-& 1.328	(0.184)	&1.049	(0.294)&	-0.390	(0.696) &	\multicolumn{4}{c}{3.139 	(0.002)}\\\hline 
\end{tabular}
\end{scriptsize}
\end{table}

%%%%%
\begin{table}[!h]
\centering \caption{\label{tab:lfsat4} Estimated parameters and joint log-likelihoods $\ell_{j|t}$ for the copula-based Markov models for ordinal time-series with covariates for satisfaction with spouse/partner $Y_5$, where a parametric copula family $C_{j|t}$ is used for the joint distribution of subsequent observations. For the best fit, the standard errors (SE) of the estimates, Wald tests ($Z$) and $p$-values are also presented.}

\begin{scriptsize}
\begin{tabular}{lcccccccc} \hline
& BVN & Frank & \multicolumn{4}{c}{Gumbel} & s.Gumbel & BVT, $\nu=3$ \\ \cline{2-9}
& Est.  &  Est.  &  Est.  & SE  & Z  & $p$-value &  Est.  &  Est. \\ \cline{2-9}
$\a_1$ & -3.75 & -4.16 & -4.09 & 0.72 & -5.65 & 0.00 & -3.19 & -3.63 \\
$\a_2$ & -3.29 & -3.62 & -3.55 & 0.72 & -4.92 & 0.00 & -2.83 & -3.20 \\
$\a_3$ & -2.90 & -3.18 & -3.12 & 0.72 & -4.33 & 0.00 & -2.53 & -2.85 \\
$\a_4$ & -2.34 & -2.59 & -2.52 & 0.72 & -3.50 & 0.00 & -2.05 & -2.33 \\
$\a_5$ & -1.47 & -1.73 & -1.64 & 0.72 & -2.28 & 0.02 & -1.20 & -1.46 \\
Age & -0.39 & -0.28 & -0.32 & 0.36 & -0.89 & 0.37 & -0.53 & -0.50 \\
Age$^2$ & 0.05 & 0.03 & 0.04 & 0.04 & 1.00 & 0.32 & 0.07 & 0.06 \\
Sex & -0.02 & 0.13 & 0.09 & 0.49 & 0.19 & 0.85 & -0.18 & -0.11 \\
Household size & -0.98 & -1.01 & -1.08 & 0.31 & -3.51 & 0.00 & -0.73 & -0.88 \\
\# of kids & -0.39 & -0.43 & -0.41 & 0.33 & -1.24 & 0.22 & -0.44 & -0.52 \\
Education &  &  &  &  &  &  &  &  \\
\hspace{2ex}No education & 0.48 & 0.52 & 0.56 & 0.13 & 4.28 & 0.00 & 0.30 & 0.45 \\
\hspace{2ex} hnd,hnc, a/o levels, cse & 0.06 & 0.09 & 0.13 & 0.09 & 1.54 & 0.12 & 0.01 & 0.08 \\
Region &  &  &  &  &  &  &  &  \\
\hspace{2ex}South & 0.18 & 0.28 & 0.18 & 0.12 & 1.49 & 0.14 & 0.06 & 0.06 \\
\hspace{2ex}Midlands & 0.13 & 0.21 & 0.11 & 0.13 & 0.85 & 0.39 & 0.04 & 0.02 \\
\hspace{2ex}North  & 0.24 & 0.27 & 0.15 & 0.12 & 1.23 & 0.22 & 0.18 & 0.12 \\
\hspace{2ex}RUK & 0.31 & 0.33 & 0.27 & 0.14 & 1.90 & 0.06 & 0.21 & 0.20 \\
Spouse Characteristics &  &  &  &  &  &  &  &  \\
\hspace{2ex} Age & -0.06 & -0.34 & -0.21 & 0.33 & -0.62 & 0.53 & 0.27 & 0.14 \\
\hspace{2ex} Age$^2$ & -0.01 & 0.02 & 0.00 & 0.03 & 0.12 & 0.91 & -0.05 & -0.03 \\
\hspace{2ex} Sex & -0.25 & -0.10 & -0.13 & 0.48 & -0.27 & 0.78 & -0.41 & -0.35 \\
\hspace{2ex}Education&  &  &  &  &  &  &  &  \\
\hspace{4ex}No education & 0.16 & 0.09 & 0.07 & 0.12 & 0.57 & 0.57 & 0.15 & 0.10 \\
\hspace{4ex}hnd,hnc, a/o levels, cse & -0.09 & -0.12 & -0.13 & 0.09 & -1.47 & 0.14 & -0.08 & -0.11 \\
\hspace{2ex}ln(Labour Income) & 0.01 & 0.01 & 0.01 & 0.01 & 1.49 & 0.14 & 0.01 & 0.01 \\
\hspace{2ex}\# Health problems & -0.14 & -0.23 & -0.18 & 0.17 & -1.02 & 0.31 & -0.08 & -0.10 \\
\hspace{2ex}Hours of wk housework & 0.02 & 0.02 & 0.02 & 0.02 & 1.00 & 0.32 & 0.01 & 0.02 \\ \hline
$\tau$ & 0.51 & 0.53 & 0.57 & 0.01 & 48.6 & 0.00 & 0.49 & 0.53 \\
$-\ell_{j|t}$ & 4007.0 & 3984.7 & \multicolumn{4}{c}{3967.3} & 4011.8 & 3968.2 \\ \hline
$z_0$ ($p$-value) &-& 1.558	(0.119) &	 \multicolumn{4}{c}{4.789	($<0.001$)}&	-0.514	(0.607)&	2.992	(0.003)\\\hline
\end{tabular}
\end{scriptsize}
\end{table}

%%%%%
\begin{table}[!h]
\centering \caption{\label{tab:lfsat5} Estimated parameters and joint log-likelihoods $\ell_{j|t}$ for the copula-based Markov models for ordinal time-series with covariates for  satisfaction with job $Y_6$, where a parametric copula family $C_{j|t}$ is used for the joint distribution of subsequent observations. For the best fit, the standard errors (SE) of the estimates, Wald tests ($Z$) and $p$-values are also presented.}

\begin{scriptsize}
\begin{tabular}{lcccccccc} \hline
 & BVN & Frank & Gumbel & s.Gumbel & \multicolumn{4}{c}{BVT, $\nu=8$} \\ \cline{2-9}
 & Est.  &  Est.  &  Est.  &  Est.  &  Est.  & SE  & Z  & $p$-value  \\ \cline{2-9}
$\a_1$ & -0.29 & -0.39 & -0.43 & -0.35 & -0.47 & 0.43 & -1.10 & 0.27 \\
$\a_2$ & 0.40 & 0.31 & 0.27 & 0.33 & 0.22 & 0.43 & 0.52 & 0.60 \\
$\a_3$ & 1.26 & 1.17 & 1.12 & 1.17 & 1.07 & 0.43 & 2.52 & 0.01 \\
$\a_4$ & 2.46 & 2.37 & 2.30 & 2.35 & 2.26 & 0.43 & 5.28 & 0.00 \\
$\a_5$ & 3.88 & 3.78 & 3.72 & 3.77 & 3.69 & 0.43 & 8.61 & 0.00 \\
Age & -0.10 & -0.13 & -0.12 & -0.17 & -0.17 & 0.20 & -0.87 & 0.39 \\
Age$^2$ & 0.02 & 0.02 & 0.02 & 0.02 & 0.02 & 0.02 & 1.09 & 0.28 \\
Sex & -0.03 & -0.05 & -0.03 & -0.03 & -0.03 & 0.04 & -0.73 & 0.46 \\
Household size & -0.58 & -0.55 & -0.67 & -0.46 & -0.55 & 0.30 & -1.85 & 0.06 \\
\# of kids & 0.41 & 0.36 & 0.49 & 0.33 & 0.40 & 0.32 & 1.26 & 0.21 \\
Education &  &  &  &  &  &  &  &  \\
\hspace{2ex}No education & -0.08 & -0.09 & -0.14 & -0.02 & -0.08 & 0.08 & -1.01 & 0.31 \\
\hspace{2ex} hnd,hnc, a/o levels, cse & -0.01 & -0.02 & -0.02 & 0.02 & 0.00 & 0.05 & 0.05 & 0.96 \\
Region &  &  &  &  &  &  &  &  \\
\hspace{2ex}South & -0.19 & -0.19 & -0.17 & -0.22 & -0.20 & 0.09 & -2.12 & 0.03 \\
\hspace{2ex}Midlands & -0.16 & -0.16 & -0.12 & -0.17 & -0.15 & 0.10 & -1.54 & 0.12 \\
\hspace{2ex}North  & -0.26 & -0.26 & -0.24 & -0.27 & -0.27 & 0.09 & -2.87 & 0.00 \\
\hspace{2ex}RUK & -0.29 & -0.29 & -0.25 & -0.31 & -0.29 & 0.11 & -2.67 & 0.01 \\
Satisfaction with job pay &  &  &  &  &  &  &  &  \\
\hspace{2ex} Medium & 0.38 & 0.38 & 0.39 & 0.39 & 0.39 & 0.05 & 7.97 & 0.00 \\
\hspace{2ex} Very & 0.74 & 0.73 & 0.74 & 0.74 & 0.74 & 0.05 & 14.3 & 0.00 \\
Satisfaction with job security &  &  &  &  &  &  &  &  \\
\hspace{2ex} Medium & 0.41 & 0.43 & 0.40 & 0.40 & 0.41 & 0.05 & 7.43 & 0.00 \\
\hspace{2ex} Very & 0.64 & 0.65 & 0.61 & 0.63 & 0.62 & 0.05 & 11.3 & 0.00 \\
Satisfaction with work itself &  &  &  &  &  &  &  &  \\
\hspace{2ex} Medium & 0.88 & 0.87 & 0.84 & 0.89 & 0.86 & 0.06 & 13.3 & 0.00 \\
\hspace{2ex} Very & 1.62 & 1.61 & 1.56 & 1.62 & 1.57 & 0.07 & 22.8 & 0.00 \\
Satisfaction with hours worked  &  &  &  &  &  &  &  &  \\
\hspace{2ex} Medium & 0.18 & 0.17 & 0.16 & 0.18 & 0.17 & 0.05 & 3.41 & 0.00 \\
\hspace{2ex} Very & 0.46 & 0.44 & 0.42 & 0.46 & 0.43 & 0.05 & 8.14 & 0.00 \\ \hline 
$\tau$ & 0.20 & 0.22 & 0.20 & 0.18 & 0.21 & 0.01 & 15.2 & 0.00 \\
$-\ell_{j|t}$ & 5245.5 & 5241.5 & 5234.8 & 5261.7 & \multicolumn{4}{c}{5228.9} \\ \hline 
$z_0$ ($p$-value) &-&0.705	(0.481)&	1.336	(0.181)	&-2.261	(0.024) &\multicolumn{4}{c}{	2.454	(0.014)}\\\hline
\end{tabular}
\end{scriptsize}
\end{table}

%%%%%
\begin{table}[!h]
\centering \caption{\label{tab:lfsat8} Estimated parameters and joint log-likelihoods $\ell_{j|t}$ for the copula-based Markov models for ordinal time-series with covariates for  satisfaction with use of leisure time $Y_7$, where a parametric copula family $C_{j|t}$ is used for the joint distribution of subsequent observations. For the best fit, the standard errors (SE) of the estimates, Wald tests ($Z$) and $p$-values are also presented.}

\begin{scriptsize}
\begin{tabular}{lcccccccc} \hline
 & BVN & Frank & Gumbel & s.Gumbel & \multicolumn{4}{c}{BVT, $\nu=4$} \\ \cline{2-9}
 & Est.  &  Est.  &  Est.  &  Est.  &  Est.  & SE  & Z  & $p$-value  \\ \cline{2-9}
$\a_1$ & -2.67 & -2.77 & -2.68 & -2.33 & -2.43 & 0.46 & -5.33 & 0.00 \\
$\a_2$ & -2.05 & -2.12 & -2.03 & -1.78 & -1.82 & 0.46 & -3.99 & 0.00 \\
$\a_3$ & -1.37 & -1.46 & -1.35 & -1.14 & -1.15 & 0.46 & -2.52 & 0.01 \\
$\a_4$ & -0.51 & -0.63 & -0.51 & -0.28 & -0.29 & 0.46 & -0.63 & 0.53 \\
$\a_5$ & 0.47 & 0.36 & 0.39 & 0.75 & 0.69 & 0.46 & 1.52 & 0.13 \\
Age & -0.58 & -0.62 & -0.49 & -0.54 & -0.47 & 0.21 & -2.23 & 0.03 \\
Age$^2$ & 0.07 & 0.08 & 0.06 & 0.07 & 0.06 & 0.02 & 2.49 & 0.01 \\
Sex & 0.10 & 0.07 & 0.05 & 0.13 & 0.10 & 0.05 & 1.85 & 0.06 \\
Household size & -0.85 & -0.88 & -0.92 & -0.71 & -0.80 & 0.28 & -2.84 & 0.00 \\
\# of kids & -0.43 & -0.45 & -0.33 & -0.50 & -0.47 & 0.30 & -1.55 & 0.12 \\
Education &  &  &  &  &  &  &  &  \\
\hspace{2ex}No education & -0.02 & 0.00 & -0.07 & 0.03 & -0.03 & 0.09 & -0.31 & 0.75 \\
\hspace{2ex} hnd,hnc, a/o levels, cse & 0.04 & 0.05 & 0.00 & 0.11 & 0.06 & 0.06 & 0.96 & 0.34 \\
Region &  &  &  &  &  &  &  &  \\
\hspace{2ex}South & 0.08 & 0.04 & 0.07 & 0.11 & 0.12 & 0.11 & 1.11 & 0.27 \\
\hspace{2ex}Midlands & 0.06 & 0.04 & 0.05 & 0.07 & 0.08 & 0.11 & 0.69 & 0.49 \\
\hspace{2ex}North  & -0.01 & -0.07 & 0.02 & -0.01 & 0.04 & 0.11 & 0.35 & 0.72 \\
\hspace{2ex}RUK & 0.09 & 0.02 & 0.10 & 0.09 & 0.14 & 0.12 & 1.14 & 0.25 \\
Frequency of leisure activities &  &  &  &  &  &  &  &  \\
\hspace{2ex}walk/swim/play sport & 0.12 & 0.11 & 0.09 & 0.11 & 0.09 & 0.03 & 2.67 & 0.01 \\
\hspace{2ex}watch live sport  & 0.14 & 0.16 & 0.12 & 0.14 & 0.12 & 0.04 & 3.30 & 0.00 \\
\hspace{2ex}cinema & -0.04 & -0.03 & -0.03 & -0.03 & -0.03 & 0.03 & -0.93 & 0.35 \\
\hspace{2ex}theatre/concert  & 0.04 & 0.03 & 0.03 & 0.04 & 0.03 & 0.03 & 1.07 & 0.28 \\
\hspace{2ex}out for a drink & 0.14 & 0.14 & 0.10 & 0.13 & 0.11 & 0.04 & 2.43 & 0.02 \\
\hspace{2ex}work in garden & -0.04 & -0.04 & -0.03 & -0.03 & -0.02 & 0.05 & -0.46 & 0.64 \\
\hspace{2ex}diy, car maintenance & -0.03 & -0.01 & -0.02 & -0.02 & -0.02 & 0.04 & -0.40 & 0.69 \\
\hspace{2ex}attend evening classes & 0.16 & 0.15 & 0.14 & 0.14 & 0.14 & 0.03 & 4.47 & 0.00 \\
\hspace{2ex}attend local groups  & 0.06 & 0.06 & 0.06 & 0.04 & 0.05 & 0.04 & 1.40 & 0.16 \\ \hline
$\tau$ & 0.36 & 0.39 & 0.39 & 0.38 & 0.38 & 0.01 & 31.52 & 0.00 \\
$-\ell_{j|t}$ & 6141.7 & 6124.7 & 6137.3 & 6113.1 & \multicolumn{4}{c}{6077.5} \\ \hline
$z_0$ ($p$-value) &1.462	(0.144)&	0.384	(0.701)&	2.189	(0.029)&	\multicolumn{4}{c}{3.999	($<0.001$)}\\ \hline
\end{tabular}
\end{scriptsize}
\end{table}

Tables \ref{tab:lfsato} - \ref{tab:lfsat8} give the estimated parameters and joint log-likelihoods $\ell_{j|t}$ for the copula-based Markov models for ordinal time-series with covariates for the seven satisfaction equations, where a parametric copula family is used for the joint distribution of subsequent observations. For the best fit, according to the likelihood principle, we also calculate standard errors (SE) and corresponding Wald tests and $p$-values. 
 SEs of  estimates have been obtained via the gradients and the Hessian computed numerically during the maximization process. Assuming that the usual regularity conditions \citep{serfling80} for asymptotic  maximum likelihood theory hold for the bivariate model as well as for its margins we have that  the estimates are asymptotically normal. Therefore we also build Wald tests to statistically judge the effect of any covariate.

\cite{liang&zeger86} noted even though parameter estimates from univariate analysis ignoring the association remain consistent,  they are inefficient. When using copula terms in the likelihood improves asymptotic efficiency over the independence estimating equations. 
\cite{Prokhorov&Schmidt2009} acknowledged that the efficiency gains will come at the expense of an asymptotic bias if the joint distribution is misspecified.  
However, our results show that  the effect of misspecifying the copula choice can be seen as minimal for both the univariate parameters and Kendall’s tau, since  (a) the univariate parameters are a univariate inference, and hence, it is the univariate marginal distribution that matters and not the type of the copula, and (b)  Kendall’s tau only accounts for the dependence dominated by the middle of the data, and it is expected to be similar amongst different families of copulas. In essence, given that tail dependence varies, the effect of different tail behaviours is  reflected in predictive inferences that depend on the joint distribution, e.g.,  the Vuong's statistic.

The best fit for the ordinal time-series, as expected, is based on BVT copulas with a small $\nu$ (according to the likelihood principle), where there is a big and statistical significant improvement over the  autoregressive-to-anything (BVN copula-based Markov) model according to Vuong's statistics. This result suggests skewness to both upper and lower tail for subsequent (in time) observations, i.e. more probability in both joint tails of the various univariate time-series. In particular, the  BVT copula, which is a radially symmetric, provides the best fit for the 6 out 7 univariate time-series. Hence,  according to  
\cite{Prokhorov&Schmidt2009}  our models are robust to the estimation of the regression parameters. Some interpretation of the estimated regression coefficients for each ordinal times-series is provided below.

In Table \ref{tab:lfsato},  generic satisfaction results suggest a drop (at an increasing rate) with age while  gender and education do not seem to matter. Smaller households are happier and with the exception of Midlands everywhere is happier than London. Improved health (i.e. fewer health problems) and higher satisfaction with job characteristics significantly improve life satisfaction. However,  absolute or relative income, house type, choice of partner or frequency of leisure activities have little direct effect on life satisfaction over and above what they might have through their respective domain satisfactions. 

As expected, in Table \ref{tab:lfsat1}  satisfaction with health goes down (at an increasing rate) with age and number of health problems, which has a large and highly significant effect. No gender or education differences are observed, while South of England and non-English regions report higher values than London.

In Table \ref{tab:lfsat2}, satisfaction with income is  significantly lower for the less educated and those in larger households but exhibits no age or gender variation. Looking at domain specific variables, with the exception of transfer and benefit income, all absolute income variables improve income well-being with the labour component exhibiting the strongest effect. Transfer income plays little role, whereas benefit income has a negative sign suggesting that those with higher benefit income are less satisfied with their financial well-being. Finally, relative position of the individual within their region (based on total annual income) bears no influence in determining domain income satisfaction.

In Table \ref{tab:lfsat3}, house satisfaction decreases (at an increasing rate) with age and size of the household, while it increases with education and house value and does not seem to change with region. Compared to a detached house all other house types result in lower satisfaction level. Similarly, compared to a very low house value, all value increments imply improvement in house domain satisfaction.

Satisfaction with spouse is the only exception in our data, with the Gumbel copula marginally providing the best fit and suggesting more probability in the upper joint tail. Overall age, gender and number of kids do not influence satisfaction with partner, whereas those less or not educated tend to be happier with the partners, as do those living in the Midlands and rest of UK compared to London. Looking at partners, little variation in domain satisfaction is explained by partners' individual  characteristics (Table \ref{tab:lfsat4}).

In Table \ref{tab:lfsat5}, none of the demographics characteristics nor education are important in explaining domain satisfaction with one's job. However, all regions report lower satisfaction compared to London, while improved satisfaction with job pay, security, work hours and work itself all significantly and strongly improve overall job satisfaction.

In Table \ref{tab:lfsat8}, satisfaction with use of leisure time goes down with age (at an increasing rate) and being lower for females and those in larger households, whereas education, more kids or geographical location have no effect. Increasing frequency of various leisure activities, on the other hand, indeed improve domain satisfaction with sports (playing or watching), out for drinks and evening classes begin the most significant.

\subsection{Fitted joint copula-based Markov models}

A joint copula-based Markov model joins the various satisfaction ordinal time-series, where the best fit  bivariate copula family from the preceding subsection  is used for the joint distribution of subsequent observations for each ordinal time-series. Since the sample is a mixture of populations (e.g., unobserved individual traits/characteristics) the MVT copula would be in theory a potential model to join the univariate time-series.

Once again, a number of different copulas are tried to form the joint distribution, i.e. the MVN and  MVT with $\nu=\{5,10,15\}$. 
Given that all together we have $197$ $\bigl($Step 1(c)$\bigr)$ $+21$ (Step 2)  parameters to be estimated, makes the third step of the estimation approach in Section  \ref{estimation} infeasible.  Hence the  estimated latent correlations, their SEs, and joint log-likelihoods $\ell_{1\ldots d|t}$ from these joint models in Table \ref{tab:corrs} are derived using  the second step of the estimation procedure. The SEs of the estimated latent correlations are obtained by the inversion of the Hessian matrix. These SEs are adequate to assess the flatness of the log-likelihood. Proper SEs that account for the estimation of all parameters can be obtained by jackknifing the two-stage estimation procedure.

\begin{table}[!h]
  \centering
  \caption{\label{tab:corrs}Estimated latent correlations, their SEs, and joint log-likelihoods $\ell_{1\ldots d|t}$ for the joint copula-based Markov models for ordinal time-series with covariates, where the best fit  bivariate copula family is used for the joint distribution of subsequent observations for each ordinal time-series, and, an MVT parametric copula family is used for the joint distribution of joint observations.}
\begin{tabular}{ccccccccc} \hline
          & \multicolumn{2}{c}{MVN} & \multicolumn{2}{c}{MVT, $\nu=5$} & \multicolumn{2}{c}{MVT, $\nu=10$} & \multicolumn{2}{c}{MVT, $\nu=15$} \\ \cline{2-9}
    	  &Est. & SE&Est. & SE&Est. & SE&Est. & SE \\ \cline{2-9}
$\rho_{12}$ & 0.381 & 0.015 & 0.365 & 0.017 & 0.379 & 0.016 & 0.383 & 0.016 \\
$\rho_{13}$ & 0.300 & 0.016 & 0.297 & 0.017 & 0.308 & 0.017 & 0.310 & 0.017 \\
$\rho_{14}$ & 0.328 & 0.016 & 0.323 & 0.018 & 0.335 & 0.017 & 0.337 & 0.017 \\
$\rho_{15}$ & 0.461 & 0.016 & 0.440 & 0.018 & 0.456 & 0.017 & 0.461 & 0.017 \\
$\rho_{16}$ & 0.357 & 0.015 & 0.361 & 0.017 & 0.369 & 0.016 & 0.369 & 0.017 \\
$\rho_{17}$ & 0.442 & 0.014 & 0.441 & 0.015 & 0.451 & 0.015 & 0.452 & 0.014 \\
$\rho_{23}$ & 0.295 & 0.016 & 0.290 & 0.017 & 0.300 & 0.017 & 0.301 & 0.016 \\
$\rho_{24}$ & 0.222 & 0.016 & 0.217 & 0.018 & 0.229 & 0.017 & 0.230 & 0.017 \\
$\rho_{25}$ & 0.191 & 0.018 & 0.173 & 0.019 & 0.188 & 0.019 & 0.192 & 0.019 \\
$\rho_{26}$ & 0.215 & 0.016 & 0.212 & 0.018 & 0.221 & 0.018 & 0.223 & 0.017 \\
$\rho_{27}$ & 0.284 & 0.015 & 0.274 & 0.017 & 0.286 & 0.017 & 0.289 & 0.016 \\
$\rho_{34}$ & 0.338 & 0.015 & 0.344 & 0.016 & 0.351 & 0.016 & 0.350 & 0.016 \\
$\rho_{35}$ & 0.184 & 0.018 & 0.173 & 0.020 & 0.186 & 0.019 & 0.189 & 0.020 \\
$\rho_{36}$ & 0.262 & 0.016 & 0.256 & 0.017 & 0.268 & 0.017 & 0.270 & 0.017 \\
$\rho_{37}$ & 0.220 & 0.016 & 0.218 & 0.018 & 0.228 & 0.017 & 0.229 & 0.017 \\
$\rho_{45}$ & 0.271 & 0.018 & 0.262 & 0.019 & 0.274 & 0.019 & 0.277 & 0.019 \\
$\rho_{46}$ & 0.172 & 0.017 & 0.178 & 0.019 & 0.185 & 0.017 & 0.185 & 0.019 \\
$\rho_{47}$ & 0.262 & 0.016 & 0.253 & 0.017 & 0.265 & 0.017 & 0.268 & 0.017 \\
$\rho_{56}$ & 0.231 & 0.019 & 0.223 & 0.019 & 0.233 & 0.020 & 0.235 & 0.019 \\
$\rho_{57}$ & 0.275 & 0.018 & 0.263 & 0.018 & 0.276 & 0.019 & 0.279 & 0.019 \\
$\rho_{67}$ & 0.188 & 0.016 & 0.190 & 0.018 & 0.198 & 0.017 & 0.199 & 0.019 \\ \hline
$-\ell_{1\ldots d|t}$ & \multicolumn{2}{c}{34950.99} & \multicolumn{2}{c}{34841.91} & \multicolumn{2}{c}{34747.89} & \multicolumn{2}{c}{34766.48} \\     \hline
Vuong's & $z_0$ & $p$-value  & $z_0$ & $p$-value & $z_0$ & $p$-value & $z_0$ & $p$-value\\
test &  \multicolumn{2}{c}{-}&2.606  & 0.009&7.634    &   $<0.001$ & 9.125      & $<0.001$\\
\hline
\end{tabular}    	  
\begin{flushleft}
\begin{scriptsize}
Subscript 1 denotes satisfaction with life overall; 2 satisfaction with health; 3 satisfaction with income; 4 satisfaction with  house/flat; 5 satisfaction with spouse/partner; 6 satisfaction with job; 7 satisfaction with use of leisure time. 
\end{scriptsize}
\end{flushleft}
\end{table}

According to the likelihood principle  the best fit is based on an MVT copula with 10 degrees of freedom. 
In this example, it is highlighted that a joint model with an MVT copula is plausible for a population that is a mixture of subpopulations, while a MVN model might be adequate for smaller homogeneous subgroups. This is confirmed by the Vuong's statistic of 7.634 (p-value $<$ 0.001) reported in the final row of Table \ref{tab:corrs}, which establishes clear superiority of the MVT over the MVN.
The fact that the best-fitting copula for the joint model is the MVT with $\nu=10$ (instead of BVN) suggests  positive tail dependence in the data, i.e. individuals reporting high satisfaction tend to do so across multiple domains, while correspondingly for those reporting low satisfaction.  
Furthermore,  a joint copula-based Markov model leads to better inferences than a copula-based Markov model with independence among the different satisfactions since the likelihood has been improved by 
$2012.0=-34747.9-(-4689.2-5725.9-5698.1-5373.0-3967.3-5228.9-6077.5)$.

Overall, all latent correlations are positive and highly statistically significant providing evidence for the multivariate nature of well-being, while suggesting that increases in life satisfaction in one domain result in further increase in satisfaction in another and the presence of potential multiplying ripple effects. The strongest latent correlations are realised between generic satisfaction and various domain satisfaction responses, i.e. Spouse, Leisure, Health and Job domain satisfaction appearing to have the strongest links to overall well-being, followed by House and lastly Income domain satisfaction. Moving on, cross-domain associations  are slightly mitigated but still various pairs emerge as important. Between domain satisfactions of income-house, health-income and health-leisure exhibit the strongest links, followed by spouse-leisure, income-job and house-leisure.

\section{\label{sec-discussion}Discussion}

In this paper, we develop a comprehensive conceptual model of life satisfaction and its constituents where a number of direct and indirect links between objective covariates and domain and generic components of well-being are captured.  Modelling dependence allows revisiting previously estimated relationships in univariate frameworks and testing their association  in a structural setting. In order to apply such structural framework, a joint copula-based Markov econometric model for ordinal time-series with covariates is developed, where each ordinal time-series is  considered a copula-based Markov model with a parametric bivariate copula for the joint distribution of subsequent observations and whose conditionals are subsequently joined through an MVT copula. We have implemented a simulated likelihood method, where the rectangles are converted to bounded integrands via the error reduction methods in \cite{genz&bretz02}, and hence the statistical efficiency of simulated likelihood is as good as maximum likelihood.

Comparing with past literature we replicate the U-shaped effect for age \citep{blanchflower_is_2008} and strong health effects \citep{dolan_we_2008}. Yet, our results fail to confirm many of the findings of past studies. For income, which has long been one of the main variables of interest driving the well-being literature, no indicator appears statistically significant for overall well-being \citep{clark_satisfaction_1996, clark_relative_2008}. Similarly for individual characteristics such as gender \citep{Alesina2004} or education \citep{blanchflower_well-being_2004}. With the exception of health and job related characteristics that appear significant in both generic and domain equations, most covariates that have been argued to directly influence overall satisfaction (i.e. housing, spouse, exercise, social life and leisure activities etc) fail to achieve significance \citep{powdthavee_putting_2008, powdthavee_i_2009, mentzakis_allowing_2011}.

This lack of direct relationships between covariates and overall well-being is suggestive of alternative underlying mechanisms that influence life satisfaction. In relation to income, past literature has put forward the importance of one's relative position and individual perception in evaluating their well-being \citep{stutzer_role_2004, clark_relative_2008, mentzakis_poor_2009}. Given the nature of domain satisfaction questions (i.e. themselves a relative measure of the well-being an individual perceives themselves as possessing in this aspect of their lives), it is mostly through them that any influence is exerted on generic satisfaction. The strong significant latent correlations confirm the presence of such links across the spectrum of domain satisfactions. The fact that job characteristics indicators (themselves satisfaction questions capturing relative position and perception) are significant in the overall well-being marginal model provides further support for this proposition.

At the extreme, our results would suggest, that covariates are less likely to directly impact generic well-being but can do so through changing individuals' satisfaction with domain well-being, which would be akin to changing individuals' relative position or perception of their status. This observation follows the rationale behind Easterlin's paradox, where once a basic level of need has been met, aspirations increase along with income with the relative position being the main aspect that continues to affect well-being \citep{easterlin_will_1995}. Such mechanism would also explain why peer and network effects have been shown to have strong effects in the literature \citep{kahneman_would_2006, clark_happy_2011}.

Looking at the literature on the structure of well-being \citep{argyle_causes_1999, van_praag_anatomy_2003}, our results empirically validate past findings with aspects of family life, social life, love life, occupational life and leisure coming up as important \citep{salvatore_appraisal_2001}, while the significance of both direct and indirect effects in the case of health lends credence to the conceptual framework and the mixture of patterns in the well-being structure \citep{brief_integrating_1993}.

Specifically, positive tail dependence suggests positive latent correlations of reported satisfaction levels over time with past high satisfaction more likely to spill-over to future periods. In other words, individuals on high satisfaction trends are more likely to continue reporting high well-being (correspondingly for those reporting low satisfaction). Positive temporal association could point to resilience or adaptation in individual happiness parallel to the``set point'' theory of happiness, in which individuals are believed to have a set happiness level that they rerun to over time after positive and negative events \citep{graham_happiness_2008, bradford_getting_2010}. Parallelly, positive temporal association could indicate that individuals of consistently high or low happiness are less likely to experience life events that will move them to the opposite happiness spectrum potentially pointing to the role of habit formation \citep{easterlin_will_1995}.

However, in the presence of shocks in domain satisfaction, overall happiness is prone to follow, something also posited within set-point theory \citep{graham_happiness_2008}. Positive tail dependence across well-being dimensions implies a drop in domain satisfaction is accompanied by corresponding changes in overall and other-domain satisfaction although the latter effects dissipate for certain domain pairs.  Looking at this from a slightly different angle, overall happiness requires happiness in all aspects of life or, alternatively, full happiness cannot be achieved without meeting a basic satisfaction level in important domains of one's life \citep{graham_happiness_2008}. The signifiant effect of health in overall well-being would suggest a complementary story to that of the latent correlations, where the former provide some type of necessary conditions for happiness that can only be realized when the latter hold.  In other words, having objectively good health improves the chances of satisfaction with life overall or maybe set the foundations for a happy life \citep{frijters_money_2004}, which however can only be enjoyed through certain personal and environmental conditions. These latter conditions that emerge from the latent correlation patterns could be usefully stylized into common clich{\'e}s, such as happiness in personal life (i.e. spousal relations) is the key to true happiness \citep{powdthavee_i_2009}, or that life is all about having fun (i.e. enjoy leisure time), or finally that income does not buy happiness, i.e., limited effect of financial satisfaction \citep{mentzakis_poor_2009}. Nevertheless, despite similar messages with past literature our results reveal substantially different underlying mechanisms that rely on the multivariate dependence component of our model that has been mostly ignored to date.

Finally, an interpretation that brings close the underlying mechanisms discussed so far in relation to the positive temporal and cross-domain dependencies would be the effect individual behavioural traits have on the evaluation of well-being. The classification of individuals according to optimistic (i.e. positively correlated high satisfaction) and pessimistic (i.e. positively correlated low satisfaction) mental predispositions and its subsequent effect on reporting patterns in subjective satisfaction questions could be the driver behind positive dependencies and offer some economic/psychological intuition behind the best fit of the MVT copula and its mixture of populations interpretation.

In conclusion, theories of generic and domain satisfaction suggest new insights can be obtained through dependence  modelling with  copulas offering a powerful and flexible tool to accommodate all necessary relationships and dependencies. 

\section*{\label{sec-asym}Appendix}
Assume a multivariate ordinal regression in which $d \geq 2$ dependent ordinal random variables $Y_{1}, \ldots, Y_{d}$ are observed together with a vector $\mathbf{x} \in \mathbb{R}^p$ of explanatory variables.
If $C(\cdot;\R)$ is the MVT copula (or any other  parametric
family of copulas) and $F_j(\cdot\; ;\; \mu,\gbf)$,
where $\mu=\x^T\bbf$ is a function of $\x$
and the $p$-dimensional regression vector $\bbf$, and $\gbf=(\alpha_1,\ldots,\alpha_{K-1})$ is the $q$-dimensional vector of the univariate cutpoints ($q=K-1$),
is a  parametric model for the $j$th univariate margin 
then
  $$C\Bigl(F_1(y_1;\mu_1,\gbf),\ldots,F_d(y_d;\mu_d,\gbf);\R\Bigr)$$
is a multivariate parametric model with univariate margins $F_1,\ldots,F_d$.

For data $\y_1,\ldots,\y_n$ and sample size $n$, 
the MVT copula model joint log-likelihood is
\begin{equation}\label{MLlik}
\ell(\bbf,\gbf,\R)\\
= \sum_{i=1}^{n}
\log{h(y_{i1},\ldots,y_{id};\bbf,\gbf,\R)},
\end{equation}
where $h(\cdot;\bbf,\gbf,\R)$ is the
joint pmf of the multivariate ordinal response vector  $\Y=(Y_{1}, \ldots, Y_{d})$, which 
can be computed through the rectangle probability: 
\begin{equation}
\label{MVNpmf}
h(\y;\bbf,\gbf,\R)=\int_{\mathcal{T}^{-1}[F_{1}(y_1-1;\mu_1,\gbf)]}^{\mathcal{T}^{-1}[F_{1}(y_1;\mu_1,\gbf)]}\cdots
\int_{\mathcal{T}^{-1}[F_{d}(y_d-1;\mu_d,\gbf)]}^{\mathcal{T}^{-1}[F_{d}(y_d;\mu_d,\gbf)]}  t_d(z_1,\ldots,z_d;\R) dz_1\ldots dz_d.
\end{equation}

In the following, we are studying the asymptotic properties of the proposed simulated likelihood for the limit (as the number of clusters increases to infinity) of the maximum simulated likelihood estimate (MSLE). We restrict ourselves to a MVN copula model with a positive  exchangeable structure, that is we took $\R$ as $(1-\rho)\I_d+ \rho \J_d$,
where $\I_d$ is the identity matrix of order $d$ and $\J_d$ is the $d\times d$ matrix of 1s.  In this special case, $d$-dimensional integrals collapse to 1-dimensional integrals \citep[p. 48]{Johnson&Kotz72} resulting in fast and accurate  MVN rectangle probabilities. Using the 1-dimensional integral method \citep{Johnson&Kotz72} to calculate rectangle MVN probabilities  (\ref{MLlik}) results in a numerically accurate likelihood method that is valid for any dimension \citep{nikoloulopoulos13b}.

By varying factors such as dimension $d$,  the amount of discreteness (number of ordinal categories), and latent correlation  for exchangeable structures,   we demonstrate patterns in the asymptotic bias of the MSLE, and assess the performance of the  simulated likelihood. 
When computing the probability limits we take a constant dimension $d$ that increases and use discrete covariates where finite number of distinct values are assumed.   Finally, without any loss of generality, we consider the case where the marginal parameters  are common to different univariate margins. 

Let the $T$ distinct cases for the ordinal response and the covariates be denoted as $$(\y^{(1)},\x^{(1)}),\ldots, (\y^{(T)},\x^{(T)}),$$ where $\y^{(t)}=(y_1^{(t)},\ldots,y_d^{(t)}),\,\x^{(t)}=(\x_1^{(t)},\ldots,\x_d^{(t)}),\, t=1,\ldots, T.$ In a random sample of size $n$, let the corresponding frequencies be denoted as $n^{(1)},\ldots, n^{(T)}$.  Assuming a probability distribution on the covariates, for $t = 1,\ldots, T$, let $p^{(t)}$ be the limit in probability of $n^{(t)}/n$ as $n\to \infty$.
For the simulated likelihood in (\ref{MLlik}), we have the limit, 
\begin{equation}\label{limitslik}
n^{-1}\ell(\bbf,\gbf,\rho)\to \sum_{t=1}^{T} p^{(t)}
\log{h(y_{1}^{(t)},\ldots,y_{d}^{(t)};\bbf,\gbf,\R)}),
\end{equation}
where $h(\y^{(t)};\bbf,\gbf,\R)$ is computed through the method proposed in \cite{genz&bretz02}. As $n\to \infty$, the limit of the MSLE, $(\bbf^{SL},\gbf^{SL},\rho^{SL})$, is the maximum of (\ref{limitslik}), while the limit of the standard maximum likelihood estimator (MLE) is the maximum of (\ref{limitslik}) where $h(\y^{(t)};\bbf,\gbf,\R)$ is computed through the 1-dimensional integral method in \cite{Johnson&Kotz72}.

We compute these limiting   MSLE in a variety of situations to show clearly if  the  SL method is good. By using these limits, we do not need Monte Carlo simulations for comparisons, and we can quickly vary parameter values and see the effects. The $p^{(t)}$ in (\ref{limitslik}) are the model based probabilities $h(\y^{(t)};\bbf,\gbf,\R)$, and computed with the 1-dimensional integral method in \cite{Johnson&Kotz72}. 

Our results are in line with the  ones in \cite{nikoloulopoulos13b} for binary and count regression with dependent data; that is the proposed MSLE are  identical with  MLE up to four decimal places.

\end{document}